\documentclass[12pt,preprint,xdvi]{aastex}

\usepackage{natbib}
\usepackage{graphicx}

\newcommand{\be}{\begin{equation}}
\newcommand{\ee}{\end{equation}}
\newcommand\eq{eq.}
\newcommand\eqs{eqs.}
\newcommand\fig{Fig.}
\newcommand\figs{Figs.}

\def\vvec{{\bf v}}
\def\xvec{{\bf x}}
\def\rvec{{\bf r}}
\def\gvec{{\bf g}}
\def\xhat{{\bf \hat{x}}}

\def\lhat{{\bf \hat{l}}}

\def\cv{{c}_{_{\rm V}}}

\def\kb{k_{_{\rm B}}}

\def\heq{H_{\rm eq}}

\begin{document}

\title{How gas-dynamic flare models powered by Petschek reconnection differ from those 
with {\em ad hoc} energy sources}

\author{D.W. Longcope}
\affil{Department of Physics, Montana State University,
  Bozeman, MT 59717}
\author{J. A. Klimchuk}
\affil{Heliophysics Division, 
NASA/GSFC, 
Greenbelt, MD  20771}

\keywords{Sun: flares}


\begin{abstract}
Aspects of solar flare dynamics, such as chromospheric evaporation and flare light-curves, have long been studied using one-dimensional models of plasma dynamics inside a static flare loop, subjected to some energy input.  While extremely successful at explaining the observed characteristics of flares, all such models so far have specified energy input {\em ad hoc}, rather than deriving it self-consistently.  There is broad consensus that flares are powered by magnetic energy released through reconnection. Recent work has generalized Petschek's basic reconnection scenario, topological change followed by field line retraction and shock heating, to permit its including into a one-dimensional flare loop model.  Here we compare the gas dynamics driven by retraction and shocking to those from more conventional static loop models energized by {\em ad hoc} source terms.  We find significant differences during the first minute, when retraction leads to larger kinetic energies and produces higher densities at the loop top, while {\em ad hoc} heating tends to rarify the loop top.  The loop-top density concentration is related to the slow magnetosonic shock, characteristic of Petschek's model, but persists beyond the retraction phase occurring in the outflow jet.  This offers an explanation for observed loop-top sources of X-ray and EUV emission, with advantages over that provided by {\em ad hoc} heating scenarios. The cooling phases of the two models are, however, notably similar to one another, suggesting observations at that stage will yield little information on the nature of energy input.  
\end{abstract}


\section{Introduction}

Our understanding of solar flares has benefitted greatly from studies using one-dimensional, gas-dynamic models.  These models pre-suppose flare plasma confined by a static magnetic field and solve for its motion parallel to the field, as well as for its thermodynamic evolution.  They solve for the evolution of of plasma within a small bundle of field lines composing a single flare {\em loop}.  The models typically add energy to the plasma through a source term representing either energy deposition by non-thermal particles, or direct heating by magnetic reconnection, through Ohmic dissipation for example.  This added energy is transported along the flare loop, ultimately heating the cooler plasma at the feet and driving chromospheric evaporation \citep{Sturrock1973,Canfield1980}.  Evaporation significantly raises the density in the hot flare loop, giving rise to the enhanced emission in EUV and X-ray which are observational signatures of a solar flare.  The subsequent cooling of the evaporated plasma dictates the evolution of the light curve at each wavelength.

One-dimensional, gas-dynamic loop models of increasing sophistication were developed throughout the early 1980s 
\citep{Antiochos1979,Nagai1980,Peres1982,Mariska1982,McClymont1983,MacNiece1984} 
and used to study the process of chromospheric evaporation \citep{Pallavicini1983,Cheng1983,Fisher1985,Fisher1985b,Fisher1985c,Emslie1985,Mariska1985,MacNiece1986}.  Restricting consideration to one-dimensional equations permitted the resolution of very small spatial scales found in the transition region, especially during a flare \citep{MacNiece1984}.  Some of the aforementioned studies invoked energy deposition by non-thermal particles primarily heating the chromosphere 
\citep{MacNiece1984,Fisher1985}.  Others used a heat source located at the loop top to represent direct effects of magnetic reconnection \citep{Cheng1983,Pallavicini1983,MacNiece1986,Gan1990}.  At least one study compared the two \citep{Emslie1985}.   Models of both kinds proved useful in interpreting spectroscopic signatures of chromospheric evaporation \citep[see][for an extensive review]{Antonucci1999}.  

Simulations of individual flaring loops have been super-posed to synthesize, and thereby understand, the light-curves of solar flares which track the cooling of the energized plasma \citep{Hori1997,Warren2005,Warren2006}.  Improvements in EUV and X-ray observations have motivated a recent revival of interest in one-dimensional, gas-dynamic modeling \citep{Bradshaw2003,Allred2005}.

Virtually all investigations have simulated loops with perfect reflectional symmetry about their apices.  The evaporated flows, driven symmetrically from each chromospheric footpoint, collide at the apex to form a stationary density concentration.  This concentration has been proposed to explain loop-top X-ray emission sometimes observed in flares \citep{Hori1997,Reeves2007}.  Since they form from chromospheric evaporation, these features will typically be somewhat cooler than the loop-top plasma had been before their formation.  It can therefore be problematic to associate them with super-hot, loop-top sources, which are typically hotter than the plasma filling the loop itself \citep{Lin1981,Lin1985,Kosugi1994,Petrosian2002,Veronig2006,Caspi2010}.

In spite of their notable successes, one-dimensional, gas-dynamic flare models have always been limited by their reliance on {\em ad hoc} heating.  According to the prevailing understanding, solar flares derive their energy from stored coronal magnetic energy released rapidly through the process of reconnection.  Since its first proposal by \citet{Sweet1958} 
and \citet{Parker1957}, there has been a steady literature devoted to studying and modeling this process 
\citep[see][for overviews]{Biskamp2000,Priest2000}.  In spite of substantial progress on this front, one-dimensional, gas-dynamic flare models continue to employ {\em ad hoc} heating not explicitly related to magnetic reconnection.  When non-thermal particles are invoked, those are given a specified energy spectrum and flux, not directly linked to a model of magnetic reconnection.  Nor does any model account for the forces from reconnection, which could affect the momentum equation of the flaring loop; the energy equation alone is modified in virtually all one-dimensional loop models.

There have been many two-dimensional and three-dimensional simulations of reconnection-powered flares \citep{Forbes1983,Magara1996,Birn2009,Karpen2012}, but the need to resolve the full coronal field, including a current sheet, compromises the resolution available for detailed loop dynamics.  Only a notable few have been able to include the field-aligned thermal conduction essential for preserving the loop-defined characteristics of the flare \citep{Yokoyama1997,Chen1999,Reeves2010}.  None of these have included a chromosphere and transition region with the same fidelity possible in one-dimensional, gas-dynamic modeling.  This naturally affects the details of the chromospheric evaporation which so many spectroscopic diagnostics probe.

The reconnection model originally proposed by \citet{Petschek1964} has been developed in a series of investigations and  proven successful in explaining fast reconnection as it might occur in a solar flare \citep{Forbes1983,Vrsnak2005}.  The key feature of this model is its assumed separation between the small length scale of the non-ideal electric field, responsible for breaking magnetic field lines, and the larger length scales on which energy is released and dissipated.  The reconnection is assumed to occur at a current sheet, embedded inside of which lies a small diffusion region.  The diffusion region defines the limited extent of the electric field which might be supported by kinetic effects \citep{Shay1998,Rogers2001,Birn2001}, anomalous resistivity \citep{Ugai1977,Magara1996,Baty2006}, or some otherwise unspecified mechanism \citep{Semenov1983,Biernat1987,Nitta2001}.  The topological change effected by this small-scale electric field initiates a larger scale outflow jet in which magnetic energy is converted first to bulk kinetic energy, and then thermalized in slow magnetosonic shocks.  The thermalization may take other forms in cases of low collisionality \citep{Gosling2005,Liu2011,Liu2012}, but most large-scale flare modeling uses fluid equations where shocks can and do occur \citep{Forbes1983,Magara1996,Yokoyama1997,Chen1999}.  

The scale separation assumed by Petschek is its most significant difference with the slower reconnection model of \citet{Sweet1958} and \citet{Parker1957} in which reconnection and energy conversion are produced by the same electric field, over the same extended layer.  Petschek-like scale separation has therefore been identified as a necessary ingredient for any magnetic reconnection mechanism to work at Alfv\'enic speeds \citep{Biskamp2001,Birn2001,Forbes2013}, which the Sweet-Parker model famously fails to do.  
Moreover, when this scale-separation does obtain, the structure on the large scales is found to approximate that of Petschek's model \citep{Erkaev2000,Biernat1987,Heyn1996}.  While the details of the diffusion region remain unclear theoretically, we hereafter use the observational fact of fast flare reconnection to posit that {\em some} localized mechanism must be at work, and that the large-scale response will therefore have a Petschek-like structure.

\begin{figure}[htp]
\epsscale{0.8}
\plotone{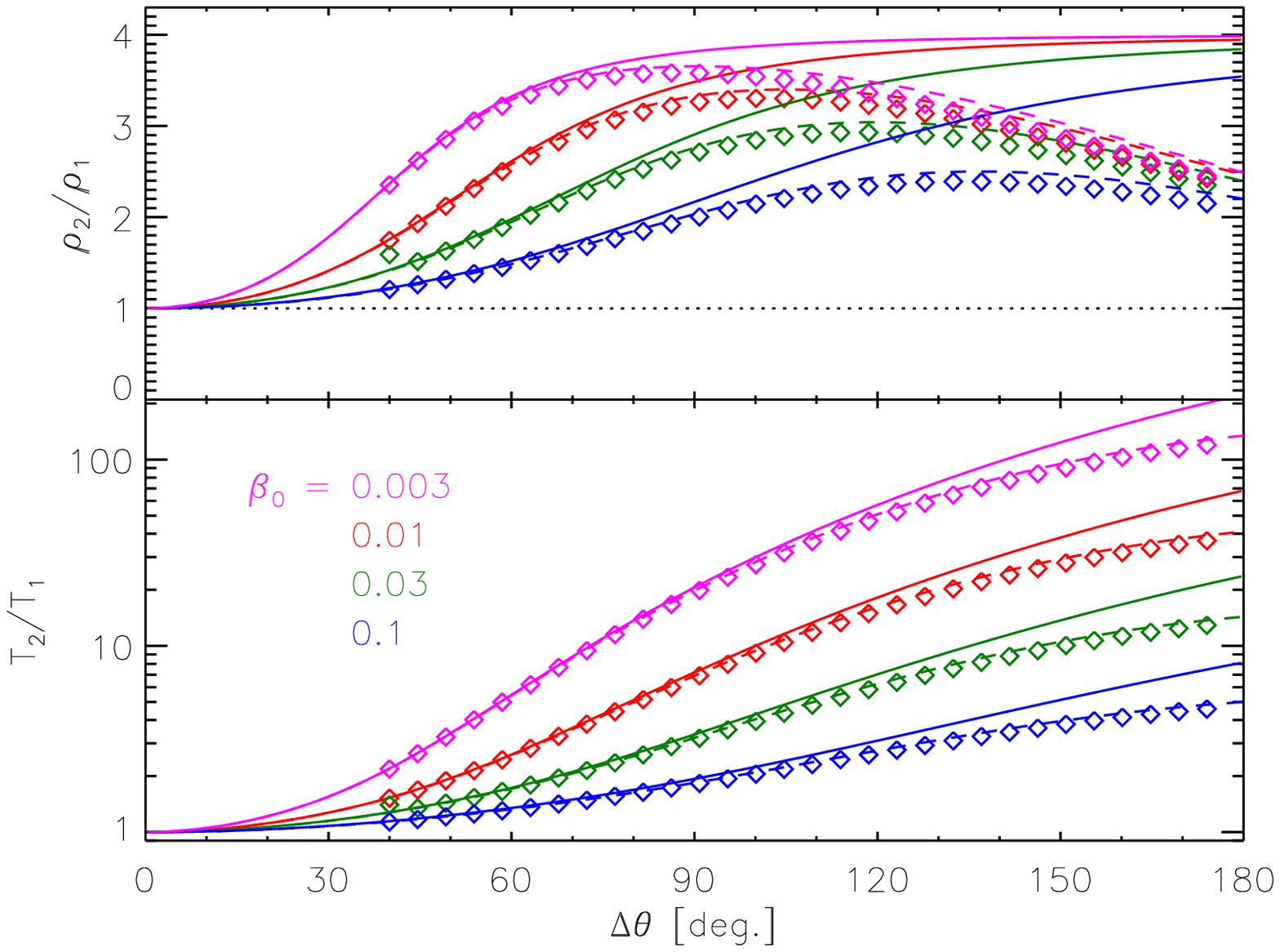}
\caption{Jumps across the slow magnetosonic shock as a function of shear angles 
$\Delta\theta$ for different models and different values of pre-reconnection plasma $\beta$:  $\beta_0=0.003$ (magenta), 0.01 (red), 0.03 (green), and 0.1 (blue).   Dashed curves show the 2.5d steady model of \citet{Soward1982b}, as reported by \citet{Forbes1989} and \citet{Vrsnak2005}.  Diamonds show the 1d Riemann problem solution of \citet{Lin1994}, and the solid curves show the TFT from \citet{Longcope2009}.  The top panel shows the ratio of post-shock density to ambient, pre-reconnection value, $\rho_2/\rho_1$; the bottom panel shows the temperature ratio, $T_2/T_1$.}
	\label{fig:dth_scan}
\end{figure}

Many of the earliest studies of Petschek's model assumed sufficient collisionality that classical thermalization could occur in slow shocks.  They further assumed these shocks were steady in their own frame and would thus satisfy Rankine-Hugoniot conditions of conservation.  This led to relations for the post-shock temperature with the shear angle across the current sheet, $\Delta\theta$, and with the plasma $\beta$ in the pre-reconnection plasma (see the dashed curves in \fig\ \ref{fig:dth_scan}).  It is noteworthy that the ratios depend little on properties of the diffusion region, such as the reconnection rate itself.

Further light was shed on the behavior of post-shock properties in later work by \citet{Lin1994}.  Instead of steady reconnection, they solved a one-dimensional Riemann problem for an initially uniform plasma whose magnetic field bent by $\Delta\theta$ at a plane.  This initial bend decomposed into a pair of fast-mode rarefaction waves, outside a pair of rotational discontinuities, outside a pair of slow magnetosonic shocks, all propagating away from the plane of initial bending. This ordering matches that of Petschek's model \citep{Petschek1967}, and when Rankine-Hugoniot relations are invoked, the post-shock temperature and density closely match those of the 2.5d reconnection models (compare diamonds to dashed lines in \fig\ \ref{fig:dth_scan}).  \citet{Lin1999b} later ran a series of time-dependent, two-dimensional MHD simulations of reconnection, and showed the outflow jets corresponded well with the 1d Riemann problem results.  A similar comparison was made by \citet{Liu2012}, using a 2.5-dimensional kinetic simulation to study non-fluid reconnection.  As described above, the solution outside the diffusion region assumed a  Petshcek-like structure, with heating (collisionless) occurring in narrow outflow jets (called ``exhaust'' flows). \citet{Liu2012} were able to match the outflow properties to the solution of a one-dimensional Riemann problem, although using collisionless variants of the traditional Rankine-Hugoniot relation.  Agreement between these disparate problems can be understood as a consequence of the scale separation invoked originally in Petschek reconnection.  The energy release in all cases occurs through a shortening of field lines on large scales --- independent of, and later than, any reconnection.

An analogous approach was taken by \citet{Linton2006} in their simplified model treating the dynamical evolution of a 
single flux tube: a {\em thin flux tube} (TFT) model.  This model considers the dynamics of post-reconnection flux retraction through the large-scale current sheet, with shear angle $\Delta\theta$.   Its basic premise is that 
the field around the current sheet is largely unaffected by the retraction of a single tube, and this external field affects the retracting flux tube only through its pressure.  Dynamics are followed beginning with the bent tube which would result from completed reconnection.  The subsequent dynamical evolution is a TFT Riemann problem in which, like its MHD counterpart  \citep{Lin1994}, the initial bend decomposes into a set of propagating shocks, surrounding a central region compressed and heated by slow shocks.  When Rankine-Hugoniot relations are invoked at the TFT shocks their properties roughly approximate those from steady Petschek  models \citep{Longcope2010,Longcope2011b}, as indicated by the solid lines in \fig\  \ref{fig:dth_scan}.  The agreement becomes increasingly poor,  especially for the density ratio, in the limit of anti-parallel fields, $\Delta\theta=180^{\circ}$.

While the original work by \citet{Linton2006} considered a distinct tube created by transient, localized reconnection (i.e.\ patchy), \fig\ \ref{fig:dth_scan} shows a resemblance to both steady and unsteady models.   The resemblance suggests, we propose, that the TFT model captures the key energetic elements of 2.5d MHD reconnection in the same way a one-dimensional Riemann model captures them \citep{Lin1999b,Liu2012}.

The present work uses the TFT model of \citet{Linton2006} as a means of introducing Petschek-like energetics into a one-dimensional flare loop model.  This is an expedient measure aimed at combining the advantages offered by one-dimensional flare loop models with an energy source more realistic than a simple {\em ad hoc} source term.  In both Petschek reconnection and the TFT model magnetic energy is released by the shortening of field lines following the topological change that occurred within the diffusion region.  In both models the retraction creates compression which occurs in shocks thereby giving rise to heating.  This combination of plasma compression with heating is absent when energy is added through an {\em ad hoc} source term in the energy equation.  Our objective is to study how the interplay of heating and compression affect observable signatures of flaring.  While we believe the TFT offers an opportunity for such a study, the quantitative discrepancies evident in \fig\ \ref{fig:dth_scan}, especially for $\Delta\theta\ga120^{\circ}$, point to  limitations of our approach.  There is, however, still value in considering a model which captures the basic elements of Petschek reconnection,  for the correct reasons, even accepting some quantitative discrepancies.

Several previous investigations have used TFT models, analytically and numerically, to study
flare dynamics \citep{Guidoni2010,Guidoni2011,Longcope2010,Longcope2010b}.
\citet{Longcope2011b} showed that the TFT model produces a hot dense concentration at the loop top which could provide an explanation for loop-top X-ray sources, alternative to colliding evaporations.   However, since none of the TFT studies performed to date have included chromospheric evaporation, those two mechanisms could not be easily compared to on another in those studies.  It is the objective of the present work to perform such modeling and to make such a comparison.

In this work we will compare the two different mechanisms for energizing flare loop models.  First we will use the TFT model to introduce energy through magnetic retraction and shocks, as in Petschek reconnection.  This energy input will be accompanied by a force in the momentum equation, the force that generates the the retraction, which will enhance plasma density near the loop top.  The retraction will generate shocks whose heating will drive evaporation, thereby enhancing the coronal density still further.  To this model we will compare the conventional, {\em ad hoc} energy input, where energy is added while the loop remains static.  Here the coronal density will be enhanced only through evaporation.  The comparison reveals, as expected, different coronal signatures during the early stages of the flare.  The differing coronal densities could also affect the radiative losses, possibly leading to different longer-term evolution.  We find, however, that this difference is very small, and it is difficult to distinguish between energization mechanisms from the loop's cooling behavior alone.

Our comparison will be presented as follows.  The next section will introduce the TFT model at the same time it reviews the more conventional static loop models.  The following section presents the result of a single TFT simulation of a retracting flux tube releasing magnetic energy and powering a flare.  Section 4 presents the result of a static loop model energized by an {\em ad hoc} heating source in its energy equation.  The source is designed to mimic the energy input of the TFT run as nearly as possible.  The results of this simulation are compared to that of the TFT model from sec.\ 3.  Finally, in section 5 we discuss the relevant points of the comparison.

\section{One-dimensional flare modeling}

\subsection{The energy equation}

One-dimensional, gas-dynamic models, including the conventional versions mentioned above and the TFT model elucidated further here, use a single coordinate $\ell$ following the magnetic field line which constitutes the loop axis.  The central element of both models is an energy equation describing the evolution of the internal energy per unit mass of the plasma's thermal (i.e.\ Maxwellian) component.   This energy, a product of specific heat, $\cv$, and temperature, $T$, changes in response to energy added to and removed from the fluid element
\begin{eqnarray}
    \cv \rho \, {dT\over dt} &=&
  - p\,\left(\lhat\cdot{\partial\vvec\over\partial\ell} ~-~ {d\ln B \over dt} \right)
  ~+~ B\, {\partial\over\partial\ell}\left( \, {\kappa\over B}\, {\partial T\over \partial\ell} \right)
  \nonumber \\[8pt]
  &~& ~-~ n_e^2\Lambda(T) 
  ~+~ \hbox{${4\over3}$}\mu \left(\lhat\cdot{\partial \vvec\over \partial\ell}\right)^2  ~+~ H(\ell,t) ~~.
  	\label{eq:erg}
\end{eqnarray}
In the equation above $\rho$, $p$ and $\vvec$ are the plasma's mass density, pressure and fluid velocity, $B$ is the magnetic field strength, and $\lhat$ is the unit vector tangent to the axis field line.  We describe below the various sources and sinks of energy which compose the right hand side (rhs) of this equation.

The first term on the rhs is adiabatic work, $-p\, dV$, and the factor in parentheses is the volumetric change in 
volume, $\rho\,d(1/\rho)/dt$, including that due to changing cross-sectional area; we derive this form below.   All remaining terms are contributions to heating, $T\, dS$.  The second term is the divergence of the heat flux.  Due to the strong magnetic field, the heat flux follows the field line very closely with a conduction coefficient given, in the classical 
Spitzer-H\"arm form, by
\be
  \kappa_{\rm sp} ~=~ \kappa_0\, T^{5/2} ~~,
  	\label{eq:kappa_sp}
\ee
with $\kappa_0=10^{-6}$ in cgs units.  This governs conduction parallel to the magnetic field and it is at least ten orders of magnitude greater than those coefficients governing perpendicular conduction.  It is this very strong field-alignment that makes flare modeling in one-dimension such a natural choice.

When temperature gradients are very large the classical heat flux, 
$\kappa_{sp}|\partial T/\partial\ell|$, would exceed the amount thermal electrons would carry if they streamed freely, 
\be
  F_{\rm fs} ~=~ \hbox{${3\over2}$}\,n_e\,\kb T\,v_{\rm th,e} ~~,
\ee
where $n_e$ is the electron number density, $v_{\rm th,e}=\sqrt{\kb T/m_e}$ is the electron thermal speed, $\kb$ is Boltzmann's constant and $m_e$ is the mass of the electron.  We keep the heat flux below some fraction of this theoretical free-streaming limit, $\xi F_{\rm fs}$, by using the modified conductivity
\be
  \kappa ~=~ \kappa_{\rm sp}\, \left[\, 1 ~+~ \left( {\kappa_{\rm sp}|\partial T/\partial\ell|\over \xi\, F_{\rm fs}}
  \,\right)^2\,\right]^{-1/2} ~~,
    	\label{eq:kappa}
\ee
in the energy equation (\ref{eq:erg}).  For simplicty we take $\xi=1$, but discuss further the effects this parameter has on the ultimate results.

The third term on the rhs of \eq\ (\ref{eq:erg}) is the energy lost from radiation.  Here we use the simplest form, optically thin losses characterized by a simple loss function $\Lambda(T)$.  While most gas-dynamics models have used a similar optically thin formulations, a few have included effects of radiative transfer, especially important in the chromosphere \citep{Fisher1985,Allred2005}.  We return to this issue below, and explain our use of the simpler form.

The fourth term is viscous heating and $\mu$ is the compressive coefficient of dynamical shear viscosity, namely the component effective along the magnetic field.  In classical theory this coefficient is intimately related to the thermal conduction and can be written
\be
  \mu(T) ~=~ {\rm Pr}\, {\kappa_{\rm sp}(T)\over \cv} ~~,
\ee
where the Prantdl number for a full ionized plasma is ${\rm Pr}\simeq 0.012$.  While this term is sometimes omitted from one-dimensional flare models owing to the fact that ${\rm Pr}\ll1$, it can be quite significant because the thermal conductivity is so great \citep{Peres1993}.  Indeed, when shocks occur it is necessary to have some form of viscosity to regularize them, and the form given here is the most physically motivated possibility.

The last term in \eq\ (\ref{eq:erg}) is an {\em ad hoc} energy source, used to stand in for sources of heat not yet understood.    One of these is the ambient heating which maintains the corona at an equilibrium temperature of one million Kelvin or more.  Another function is to replace physics which actually thermalizes energy in a flare.  We use the term only for the former purpose in our TFT simulations.  We then modify it to achieve both ends in the 
{\em ad hoc}  simulations, to which we compare.

\subsection{Petschek reconnection and the Thin Flux Tube model}

The basic problem facing any flare reconnection model, including that of \citet{Petschek1964}, is tapping some of the energy stored in the coronal magnetic field.  This energy can be formally rewritten as an integral over all the flux tubes composing the coronal volume
\be
  W_{M} ~=~ {1\over8\pi}\int\, B^2\, dV ~=~ \int\, d\Phi\, \left(\, {1\over8\pi}\int\, B\, d\ell\, \right) ~~,
  	\label{eq:WM}
\ee
where the integral inside parentheses runs along the axis of an infinitesimal flux tube (it is the energy of a field line), while the outer integral is over all flux tubes.  It is a basic tenet of the models that a flare does not add or remove magnetic flux, but simply rearranges it in order to reduce $W_M$.  This means that reconnection must produce an overall decrease of the inner integrals.  Petschek's model achieves the reduction primarily by reducing the length of each field line through the retraction which follows the topological change in the diffusion region \citep{Longcope2010}.  Even for the case of anti-parallel reconnection, where switch-off shocks dramatically reduce the field strength, the field line is ultimately ejected from the outflow with field strength back around the ambient level, but greatly reduced in length.  

In order for a small diffusion region to initiate 
a significant length change it must overlap field lines with very different global properties.  This requirement seems to demand that reconnection occur within a current sheet, or thin current layer, separating field lines whose direction differs by some finite angle $\Delta\theta$.  It is for this reason that all models of fast reconnection pre-suppose a current sheet.  We do the same here.

The present work will will use the TFT model of \citet{Linton2006} and \citet{Longcope2009} to mimic the effects of
Petschek reconnection in a one-dimensional flare loop.  The model describes the evolution of a tube with axis $\rvec(\ell,t)$ confined to move within an equilibrium current sheet.  The field strength on either side of the sheet is a fixed function of position, $B(\xvec)$, independent of time, and unaffected by the flux tube evolution.  It is further assumed that pressure balance across the tube is dominated by magnetic pressure (i.e.\ $\beta\ll1$) and thus the internal field strength matches $B(\rvec)$.  The potential energy, per unit flux, of such a tube
\be
  {\cal V}[\rvec(\ell)] ~=~ \int\, \left[ {B^2(\rvec)\over4\pi} ~+~ \hbox{${3\over 2}$}\,p \,- \rho\, \gvec\cdot\rvec(\ell)
  \, \right]\, {d\ell\over B(\rvec)} ~~,
  	\label{eq:pot_erg}
\ee
includes magnetic, thermal and gravitational potential energy, with $\gvec$ being the downward gravitational acceleration.  The differential, $d\ell/B$, is the element of volume per unit flux.  This potential energy includes the work done by the confining magnetic field on the tube \citep{Longcope2011b}, which appears to double the magnetic contribution given by \eq\ (\ref{eq:WM}).

The plasma composing the tube moves at velocity $\vvec=d\rvec/dt$, of which only the component perpendicular to the 
tangent vector, $\lhat=\partial\rvec/\partial\ell$, moves the axis.  The plasma velocity is changed by forces acting on the plasma.   The conservative forces can be derived by formal variation of the potential energy (\ref{eq:pot_erg}), assuming virtual variations in $p$ to be adiabatic with $\gamma=5/3$.  Adding a non-conservative viscous force yields the tube's momentum equation
\begin{eqnarray}
  \rho{d\vvec\over dt} &=& \left( {B^2\over 4\pi} - p \right)\, {\partial\lhat\over\partial\ell} 
  ~-~ \left( 1 +{4\pi p\over B^2}\right)\,\nabla_{\perp}\left({B^2\over 8\pi}\right)
  ~-~ \lhat\, {\partial p\over \partial \ell} 
  ~+~ \rho\,\gvec
  \nonumber \\[9pt]
  &~&  ~+~ \hbox{${4\over3}$}\,B\, {\partial\over\partial\ell}\left[ \, \lhat\,{\mu\over B}\, \left(\lhat\cdot
  {\partial \vvec\over \partial\ell}\right)\, \right] ~~,
  	\label{eq:mom}
\end{eqnarray}
where $\nabla_\perp$ is the component of the gradient perpendicular to $\lhat$.  The first two terms on the rhs, both perpendicular to the tube's axis, are primarily the Lorentz force.  The first of them, the tension force, is along the tube's curvature vector $\partial\lhat/\partial\ell=\partial^2\rvec/\partial\ell^2$, provided 
$\beta<2$.  This means it acts to straighten the tube's axis, thereby shortening the tube and
driving its retraction following reconnection.  The second magnetic force arises from the confining external field.  Any variation in its strength tends to squeeze the tube like a seed between slippery fingers.

Derivation from a potential guarantees energy conservation, even in cases where $\beta$ becomes appreciable.  
If we omit the viscous force and use the adiabatic gas law, then the total energy, combining potential (i.e.\ \eq\ [\ref{eq:pot_erg}]), and kinetic, would be strictly conserved by evolution under \eq\ (\ref{eq:mom}).  The form of the viscous term in \eq\ (\ref{eq:mom}) is such that it will decrease but never increase this total energy.  The viscous heating contribution to \eq\ (\ref{eq:erg}) exactly cancels this decrease, restoring energy conservation.  (The similarity of the terms is clear, even to the factor of $4/3$.)  It is a further consequence of the variational derivation, which by itself makes no assumptions about $\beta$, that the first two terms on the rhs of \eq\ (\ref{eq:mom}), differ from the corresponding terms in \citet{Guidoni2011} at the level of $\sim\beta$.  In that previous work the factor multiplying each term included only its leading order term, and therefore energy was conserved only up to order $\beta$.

By tracking the motion of the flux tube we tacitly assume an ideal induction equation.  There is no Ohmic diffusion or non-ideal electric field in the TFT model.  The model is therefore restricted to the dynamics {\em after} any non-ideal processes have occurred in the diffusion region.  We assume those processes are completed, have created the new flux tube (the initial condition of our model), and we ignore any heating that might have accompanied them.  Doing so yields results similar to those of standard the Petschek model because that model also ignores the energetic contribution of Ohmic diffusion in the face of the much greater contribution of post-reconnection retraction.  The tension force converts magnetic energy to bulk kinetic energy of the outflow (i.e.\ the retraction).  Shocks form and in them viscosity converts bulk kinetic energy to thermal energy through the penultimate term in \eq\ (\ref{eq:erg}).  Like Petschek's model, the TFT model converts magnetic energy to heat without recourse to Joule heating or the {\em ad hoc} heating term $H$.

Equations (\ref{eq:erg}) and (\ref{eq:mom}) are the core of the TFT.  A tube element  with mass per flux $\delta m$ will have mass density
\be
  \rho(\ell,t) ~=~ B[\rvec(\ell,t)]\, {\delta m\over\delta\ell} ~~,
  	\label{eq:rho_eq}
\ee
when extended to axial length $\delta\ell$.  This is how density is computed when following Lagrangian tube segments of constant $\delta m$, as we do below.  If one wanted to exhibit an analogue of a conventional continuity equation, one would take the time derivative of \eq\ (\ref{eq:rho_eq}), perform some algebra and obtain
\be
  {d\rho\over dt} ~=~ \rho\, {d(\ln B)\over dt} ~-~ \rho\, \lhat\cdot {\partial\vvec\over\partial\ell} ~~.
\ee
(This is the form used in the adiabatic work term in the energy equation [\ref{eq:erg}].)  Finally, the pressure 
is found from the temperature and mass density, $p=(\kb/\bar{m})\rho T$, where $\bar{m}$ is the mean mass per particle.

\section{Reconnection simulation}

\subsection{The numerical code: PREFT}

Equations (\ref{eq:erg}), (\ref{eq:mom}), and (\ref{eq:rho_eq}) are solved on a Lagrangian grid by a
code called PREFT,\footnote{PREFT stands for Post-Reconnection Evolution of a Flux Tube.}
an extension of the DEFT code of \citet{Guidoni2010}.  The flux tube is represented using discrete 
segments, or cells, with fixed mass per unit flux, $\delta m$.  The vertices joining segments move at velocity $\vvec$ which is advanced explicitly according to \eq\ (\ref{eq:mom}).  The temperature within each cell is advanced semi-implicitly 
according to \eq\ (\ref{eq:erg}); the thermal conduction term is differenced implicitly while all others are explicit.

The main advance of PREFT over DEFT is its inclusion of a crude chromosphere at the feet of the retracting flux tube.  The  chromosphere serves as a reservoir of cool material whose evaporation permits the coronal simulation to continue well after the tube retraction is completed.  With this limited goal in mind we simplify the physics by assuming a fully ionized plasma with coronal abundances, even in the chromosphere.  We therefore set the mean particle mass to $\bar{m} = 0.593\, m_p$, and the electron number density to $n_e=0.874(\rho/m_p)$, where $m_p$ is the proton mass.  The corresponding specific heat
\be
  \cv ~=~ {3\over 2}{\kb \over \bar{m}} ~~,
  	\label{eq:cv}
\ee
excludes any contribution from ionization, even as chromospheric material is heated to coronal temperatures.  The effects on evaporation of this simplifying assumption were studied in \citet{Longcope2014b}, and found to be very minor.

It is in the same spirit that we opt for an optically thin treatment of radiative losses.  While radiative transfer would be necessary to more accurately model the chromosphere, we wish only to provide a reservoir from which evaporation might be driven.
Our radiative loss function $\Lambda(T)$ is a piece-wise power-law fit to the output of {\sc Chianti} 7.1 with coronal abundances  and default ionization equilibria \citep{Landi2013,Dere1997}. 
We make this function monotonic, $\Lambda\sim T^{1.66}$, over the range $10,100\,{\rm K}<T<78,700\,{\rm K}$, omitting a peak around $20,000$ K, due to silicon, since it would make the lowest temperature regions susceptible to radiative instabilities.  For $T<10,100$ K, which is the chromosphere, we set $\Lambda=0$.

We use gravity only to stratify the chromosphere so that its high pressure might confine the flare corona.  We thus set
$g=274\,{\rm m/s^2}$ within the ends of the loop, and $g=0$ everywhere else.  In keeping with our aim of providing a simple, cool reservoir we use the full solar value of $g$, thereby assuming that small region to be effectively vertical, even if the axis is actually not.  

\begin{figure}[htp]
\epsscale{1.1}
\plotone{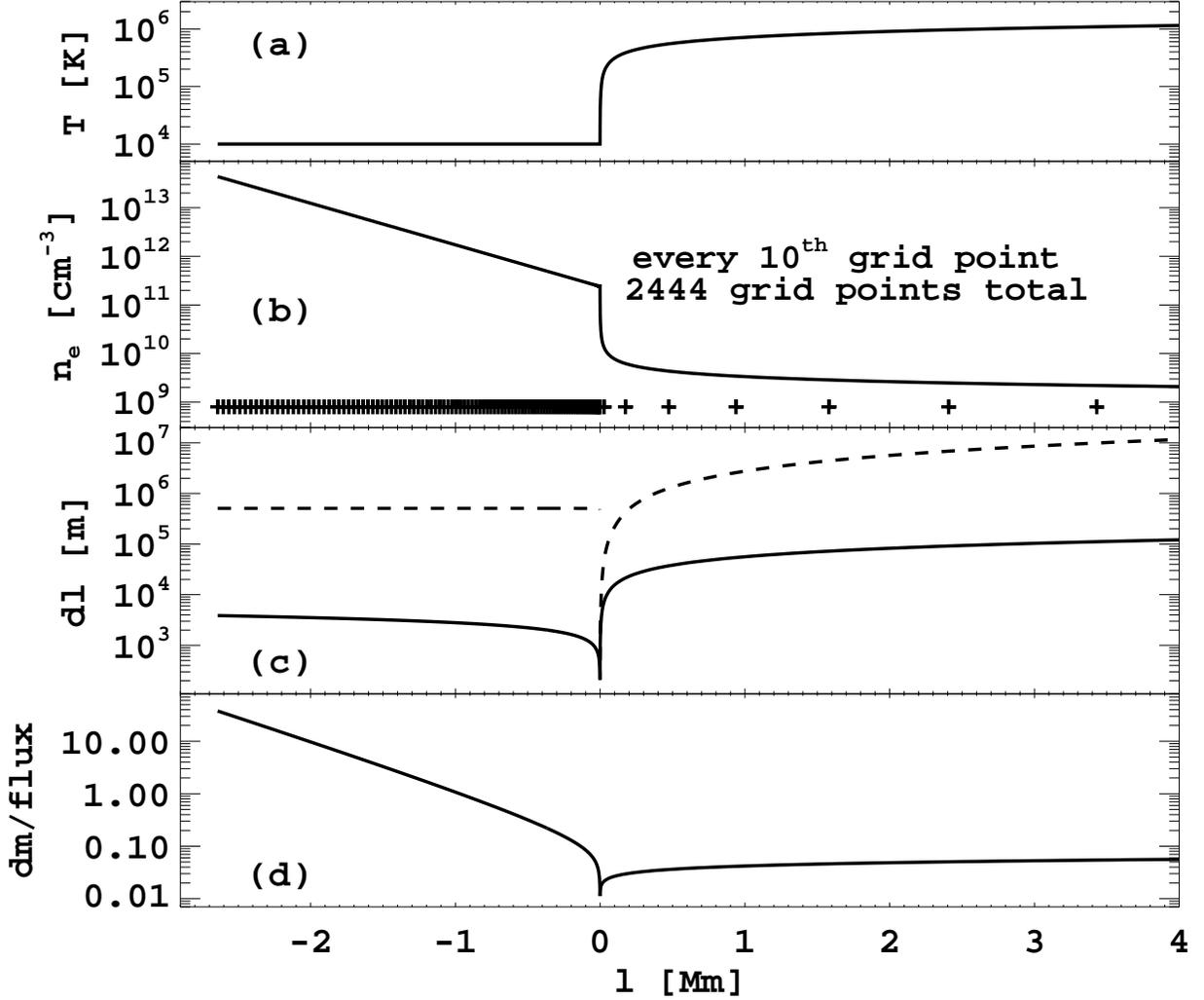}
\caption{One end of an initial equilibrium flux tube with peak temperature $T_{\rm max,0}=2$ MK and total coronal length $L_0=69.2$ Mm, of which $6.6$ Mm are shown.   An isothermal chromosphere, $T_{\rm min,0}=10,000$ K, runs 5.2 scale heights leftward, from $\ell=0$ to $\ell=-2.6$ Mm.  Panels (a) and (b) show $T(\ell)$ and $n_e(\ell)$, respectively.  Crosses in (b) show every $10^{\rm th}$ grid point.  Panel (c) shows the length of each cell, $\delta\ell$ (solid) and the local 
density scale length, $\rho/(d\rho/d\ell)$ (dashed), in units of meters.  Panel (d) shows the mass per magnetic 
flux, $\delta m$, of each cell in units of $10^{-8}\,{\rm gm/Mx}$.}
	\label{fig:grid}
\end{figure}

We initialize the coronal portion of the flux tube, where $g=0$, with an isobaric equilibrium whose temperature rises from $T_{\rm min,0}=10^4$ K to $T=T_{\rm max,0}$ over a half-length $L_0/2$.  This requires a uniform volumetric heating $\heq$ whose value, along with the pressure, is determined by $L_0$ and $T_{\rm max,0}$.  To this isobaric equilibrium we append isothermal chromospheres, $T=T_{\rm min,0}$, with no heating or radiative losses ($T_{\rm min,0}<10,100$ K) and stratified by gravity.  Figure \ref{fig:grid} shows one end of an equilibrium.

The equilibrium tube is represented on a grid whose points are distributed to resolve it.  Cells are adjusted 
to keep their mass, $\delta m$, relatively uniform above the chromosphere, as shown in \fig\ \ref{fig:grid}d.  As a result cell lengths, $\delta\ell$ (solid curve in \fig\ \ref{fig:grid}c), shrink dramatically in the transition region and chromosphere, becoming as small as $\delta\ell=209$ m at the base of the transition region, in this particular example.  This keeps them more than an order of magnitude smaller than the local density scale height (dashed curve in \fig\ \ref{fig:grid}c).  Cell masses are allowed to rise through the chromosphere, until the cell length is $\delta\ell=3.9$ km at the end of tube --- two orders of magnitude smaller than the scale height.  The upshot, summarized in table \ref{tab:IC}, is a grid where $84\%$ of the cells are packed into chromospheres composing $7\%$ of the tube's length.

\begin{table}[htbp]
\centering
\begin{tabular}{lc|c|c|}
\multicolumn{2}{l}{\sc complete tube\dotfill} & TFT run & {\em ad hoc} run\\ \hline
$N$ & & 2444 & 2766 \\
$B$ & G & 75 & 75 \\
initial full length & Mm & $74.5$ & $61.2$ \\
final full length & Mm & $61.4$ & $61.2$ \\
initial minium $\delta\ell$ & m & 209 & 168 \\[9pt]
\multicolumn{2}{l|}{\sc initial corona\dotfill} & & \\ \hline
$N$ coronal & & 397 & 397 \\
full length & Mm & $69.2$ & $55.9$ \\
$T_{\rm max,0}$ & K & $2.0\times10^6$ & $2.0\times10^6$ \\
$p$ & ${\rm erg\,cm^{-3}}$ & 0.56 & 0.69 \\
$n_e$ at apex & ${\rm cm}^{-3}$ & $1.0\times10^9$ & $1.3\times10^9$ \\
electron column of half-loop & ${\rm cm^{-2}}$ & $4.9\times10^{18}$ & $4.8\times10^{18}$ \\
$H_{\rm eq}$ & ${\rm erg\, cm^{-3}\, s^{-1}}$ & $1.0\times10^{-3}$ & $1.5\times10^{-3}$ \\[9pt]
\multicolumn{2}{l|}{\sc initial chromosphere\dotfill} & & \\ \hline
$T_{\rm min,0}$ & K & $1.0\times10^4$ & $1.0\times10^4$ \\
depth & Mm & 2.6 & 2.6 \\
$n_e$ at top & ${\rm cm}^{-3}$ & $2.1\times10^{11}$ & $2.6\times10^{11}$ \\
$n_e$ at base & ${\rm cm}^{-3}$ & $3.8\times10^{13}$ & $4.7\times10^{13}$ \\
total electron column & ${\rm cm^{-2}}$ & $1.9\times10^{21}$ & $2.4\times10^{21}$ \\
number of scale heights & & 5.2 & 5.2 \\
maximum $\delta\ell$ & km & 3.9 & 3.3
\end{tabular}
\caption{Properties of the numerical simulations}
	\label{tab:IC}
\end{table}

\subsection{The retraction simulation}

\begin{figure}[htp]
\epsscale{0.5}
\plotone{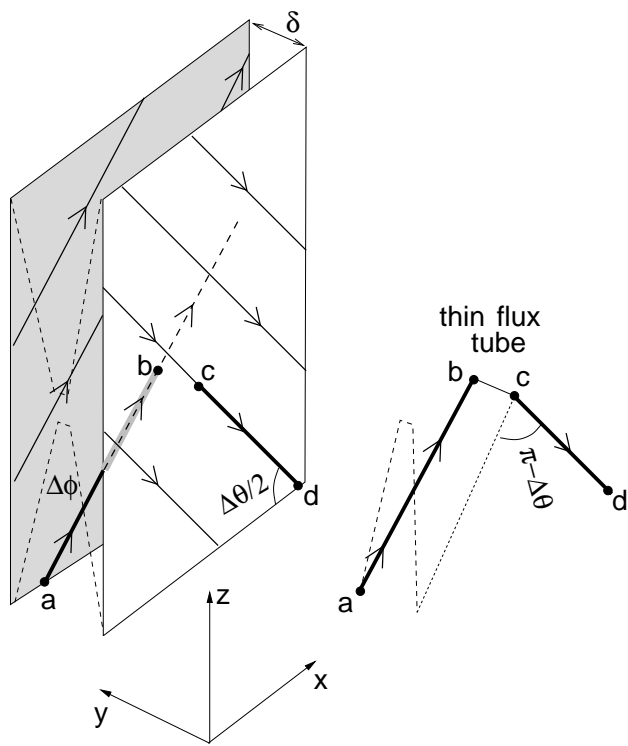}
\caption{The geometry of the current sheet housing the TFT model.  Two layers of uniform magnetic field (solid arrows), whose direction differs by shear angle $\Delta\theta$, are separated by a current sheet of thickness $\delta$.  Reconnection forms a link between point {\bf b} on the far layer and point {\bf c} on the near layer.  The result is the bent thin flux tube {\bf a}--{\bf b}--{\bf c}--{\bf d}, repeated in isolation to the right.  If the reconnected tube is viewed edge-on it could be seen to make an angle $\Delta\phi$, shown with thin dotted lines along the edge of the current sheet.  This is the reconnection jet opening angle often referred to in studies of steady Petschek reconnection.  Here we take $\delta\to0$ and thus $\Delta\phi\to0$, and work only with the shear angle $\Delta\theta$ between the external flux layers.  In this limit the initial TFT lies in the $x$--$z$ plane with  a simple bend of angle $\pi-\Delta\theta$, as shown in the right image.}
	\label{fig:geom}
\end{figure}

We simulate  flux tube retraction in a manner similar to previous investigations 
\citep{Guidoni2010,Guidoni2011,Longcope2011b}.  For simplicity, we use a current sheet separating layers of uniform magnetic field with identical magnitude $B(\xvec)=B_0=75$ G, but differing in direction by an angle $\Delta\theta$, often called the shear angle.  We take the sheet to lie in the $x$--$z$ plane (see \fig\ \ref{fig:geom}), with current flowing in the $\xhat$ direction,\footnote{The magnetic field component parallel to the current, sometimes called the {\em guide field},  is $B_x=B_0\cos(\Delta\theta/2)$.} though that current plays no role in \eqs\ (\ref{eq:erg}), (\ref{eq:mom}), or (\ref{eq:rho_eq}).  The tube is restricted to remain within the current sheet, i.e.\ in the $x$--$z$ plane.  It is worth noting that plotting the field line in the $x$--$z$ plane provides {\em face-on-view} of the sheet, rather than the more traditional {\em edge-on view} which would appear as a vertical line in the current model.\footnote{In steady Petschek models the sheet is thickened to a wedge with small opening angle, $\Delta\phi$, related to the steady reconnection electric field within the diffusion region.  This opening angle is illustrated in \fig\ \ref{fig:geom}, by giving the current sheet a finite thickness, $\delta$.  Both the sheet thickness, $\delta$, and opening angle, $\Delta\phi$, are irrelevant to the TFT model and are taken to zero with no consequences.  We do not refer to this angle, or the end-on-view, in the body work but return in the Discussion section to consider the possible effects of this simplifying assumption.}

\begin{figure}[htp]
\epsscale{1.0}
\plotone{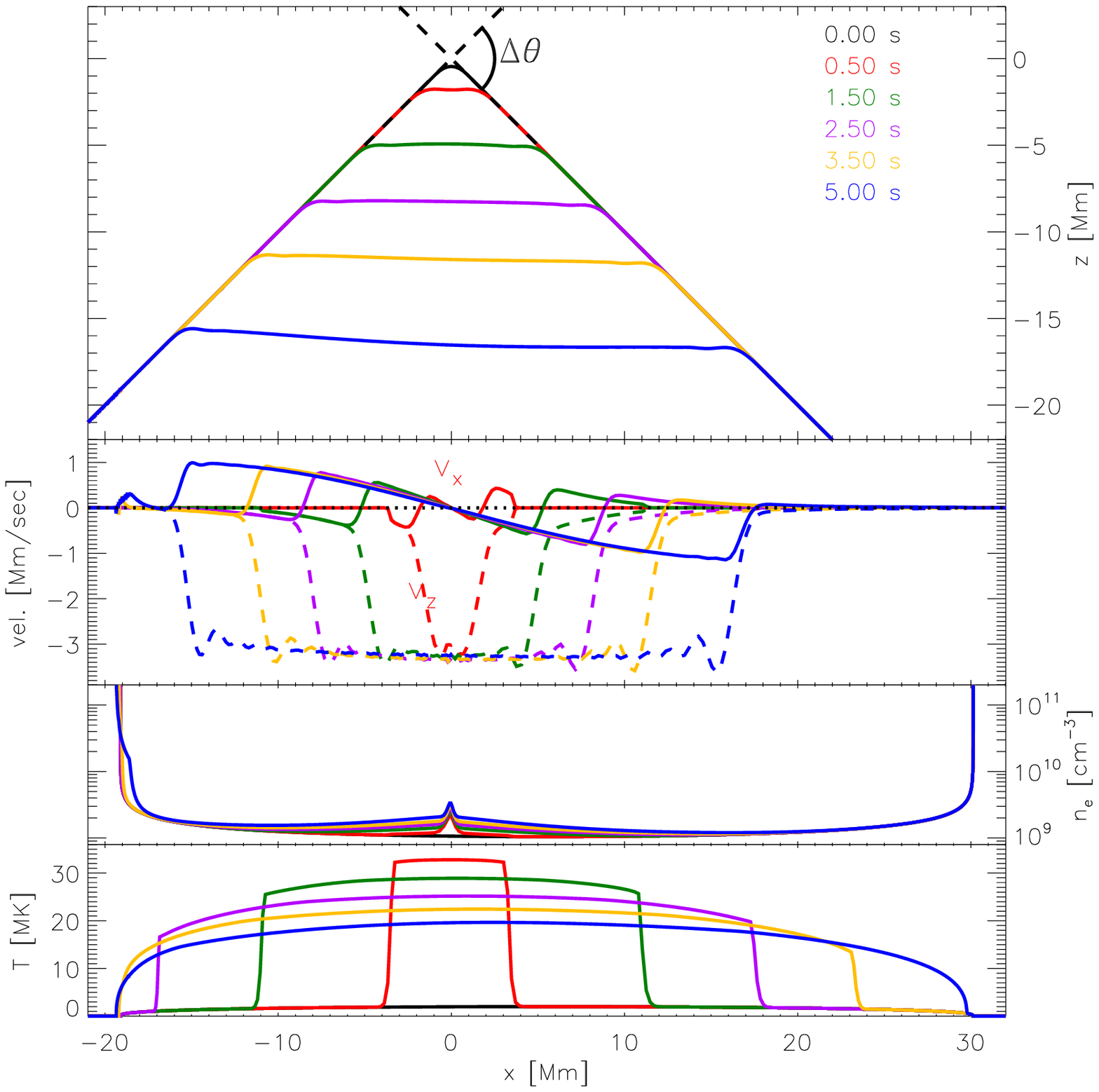}
\caption{The initial evolution of the flux tube retraction under the TFT model.  The top panel shows the tube 
axis, $\rvec(\ell,t)$ at $t=0$ (black) and 5 subsequent times, in a color code given along the right.  The reconnection at point $\xvec=(0,0)$, creates a bend at angle $\Delta\theta=90^{\circ}$, labelled.  The left footpoint, at $x=z=-20$ Mm is visible, but the right one is not.  The second panel shows $v_x$ (solid) and $v_z$ (dashed) using the same color code, plotted against horizontal coordinate $x$.  The panels below that show $n_e$, on a logarithmic scale, and $T$ on a linear scale (bottom), both plotted against $x$.}
	\label{fig:bend45_retract}
\end{figure}

For a uniform field, $B(\xvec)=B_0$, the equilibrium solution to \eq\ (\ref{eq:mom}) is a straight tube, $\partial\lhat/\partial\ell=0$.  We simulate the effect of completed reconnection by joining two straight segments at an angle 
$\Delta\theta=90^{\circ}$ --- in other words we bend the equilibrium flux tube described in the previous section by 
$180^{\circ}-\Delta\theta=90^{\circ}$ (see \fig\ \ref{fig:geom}).  Previous investigations have placed the reconnection site at the precise center of the tube, introducing a symmetry which seems unlikely in nature.  Here we choose to break this unlikely symmetry by placing the reconnection site to one side.   Beginning with a tube of coronal length, $L_0=69.2$ Mm (one of whose ends is shown in \fig\ \ref{fig:grid}), we introduce a $90^{\circ}$ bend at a point $27.0$ Mm from the left chromosphere, and $42.2$ Mm from the right (see \fig\ \ref{fig:bend45_retract}).  To help distinguish parallel from perpendicular flows, we orient the bent tube with legs at $\pm45^{\circ}$ from the $x$ axis (recall that $\gvec$ is oriented along the tube regardless of the axis orientation).  Properties of this initial tube are summarized in table \ref{tab:IC}.

Beginning from this bent initial condition we solve \eqs\ (\ref{eq:erg}), (\ref{eq:mom}), and (\ref{eq:rho_eq}).  The heating term $H$ is maintained at the very small value, $H_{\rm eq}=10^{-3}\,{\rm erg\,cm^{-3}\,s^{-1}}$, needed to sustain the initial equilibrium; it is {\em not} the source of the flare energy.  The initial evolution of the retracting tube, shown in 
\fig\ \ref{fig:bend45_retract}, is similar to those seen in previous investigations 
\citep{Guidoni2010,Guidoni2011,Longcope2011b}.  The $\Delta\theta=90^{\circ}$ bend decomposes into 
two $\Delta\theta/2=45^{\circ}$ bends,\footnote{Note that $\Delta\theta/2=45^{\circ}$ is the change in the tangent vector 
$\lhat$ across the RD.  This change makes the flux tube axis bend by an angle $180^{\circ}-\Delta\theta/2=135^{\circ}$.}
called {\em rotational discontinuities} \citep[RDs,][]{Longcope2009}, 
which move along the straight legs at the Alfv\'en speed, $v_{\rm A}\simeq4.7$ Mm/s. 
The plasma downstream of the RDs is moving downward at $v_z=v_{\rm A}\sin(\Delta\theta/2)\simeq3.3$ 
Mm/s, and inward at $v_x=\pm2v_{\rm A}\sin^2(\Delta\theta/4)\simeq\pm1.4$ Mm/s.  (The latter theoretical
value is partly offset by outward flows generated by the thermal conduction fronts.)  This downward-inward flow is directed along $\hbox{atan}(v_z/v_x)=90^{\circ}-\Delta\theta/4$, the bisector of the $135^{\circ}$ bend \citep{Longcope2009}, which is the direction of the tension force there.  

The vertical motion, the analog of the reconnection outflow, moves an approximately horizontal section of tube downward.  The horizontal motion produces a collision at the center ($x\simeq 0$) analogous to the the pair of slow mode shocks characteristic of Petschek reconnection.  
Due to the viscosity, significantly enhanced at $>20$ MK, the shocks are replaced by a smooth reversal in $v_x$ with a density plug in the middle.  The viscous heating from this central feature raises the peak temperature above 34 MK.

Thermal conduction rapidly spreads the centrally-generated heat outward in thermal conduction fronts.    Because of the flux limiter, \eq\ (\ref{eq:kappa}), the fronts become steep temperature jumps, at which the heat flux achieves its maximum value $F = \xi F_{\rm fs}\simeq1.2\times10^{10}\,{\rm erg\,cm^{-2}\,s^{-1}}$.  These jumps propagate outward at
\be
  v_{\rm fr} ~=~ \xi^{2/3}\,f_e\,\left(\, {2\over 3}\,{F\over m_e\,n_e}\,\right)^{1/3} ~\simeq~9\, {\rm Mm/s} ~~,
\ee
comparable to the post-jump electron thermal speed, where $f_e=0.52$ is the fraction of particles accounted for by electrons.  This exceeds the Alfv\'en speed, so the conduction fronts move out ahead of the RDs.

A novel feature of this simulation is that the conduction fronts reach the chromosphere and drive chromospheric evaporation.  This occurs first at the nearer (left) footpoint, just before $t=3.5$ sec.  By $t=5.0$ sec a small positive bump in $v_x$, to the left of the RD, is evidence of the evaporation flow in \fig\ \ref{fig:bend45_retract}.  The right front appears to have just reached the farther chromosphere at $t=5.0$ sec.

We halt the tube's retraction at $t=5.0$ sec when its total length has decreased to $61.4$ Mm, from its original $74.5$ 
(see table \ref{tab:IC}).  We do this by artificially straightening the tube in an effort to mimic the effect of the tube reaching the base of the current sheet \citep{Guidoni2011}, even though our uniform current sheet has no actual base.  Were there an actual base, the horizontal field beneath it, the post-flare arcade, would exert an upward force to bring the retracting tube to rest.  If the downward flow speed were faster than the fast magnetosonic speed this might form a fast mode termination shock \citep[FMTS,][]{Forbes1983,Forbes1986c}, otherwise it would launch a fast mode wave.  In either event the tube's perpendicular velocity would be eliminated with little effect on the parallel velocity \citep{Guidoni2011}.  We mimic this effect by making the tube axis lie long the $x$ axis, and setting $v_x(x)$ equal to $\lhat\cdot\vvec$ at the instant before straightening.  We then continue solving the TFT equations, (\ref{eq:erg}), (\ref{eq:mom}) and (\ref{eq:rho_eq}), although now they include only those terms from traditional static, one-dimensional models.

\begin{figure}[htp]
\plotone{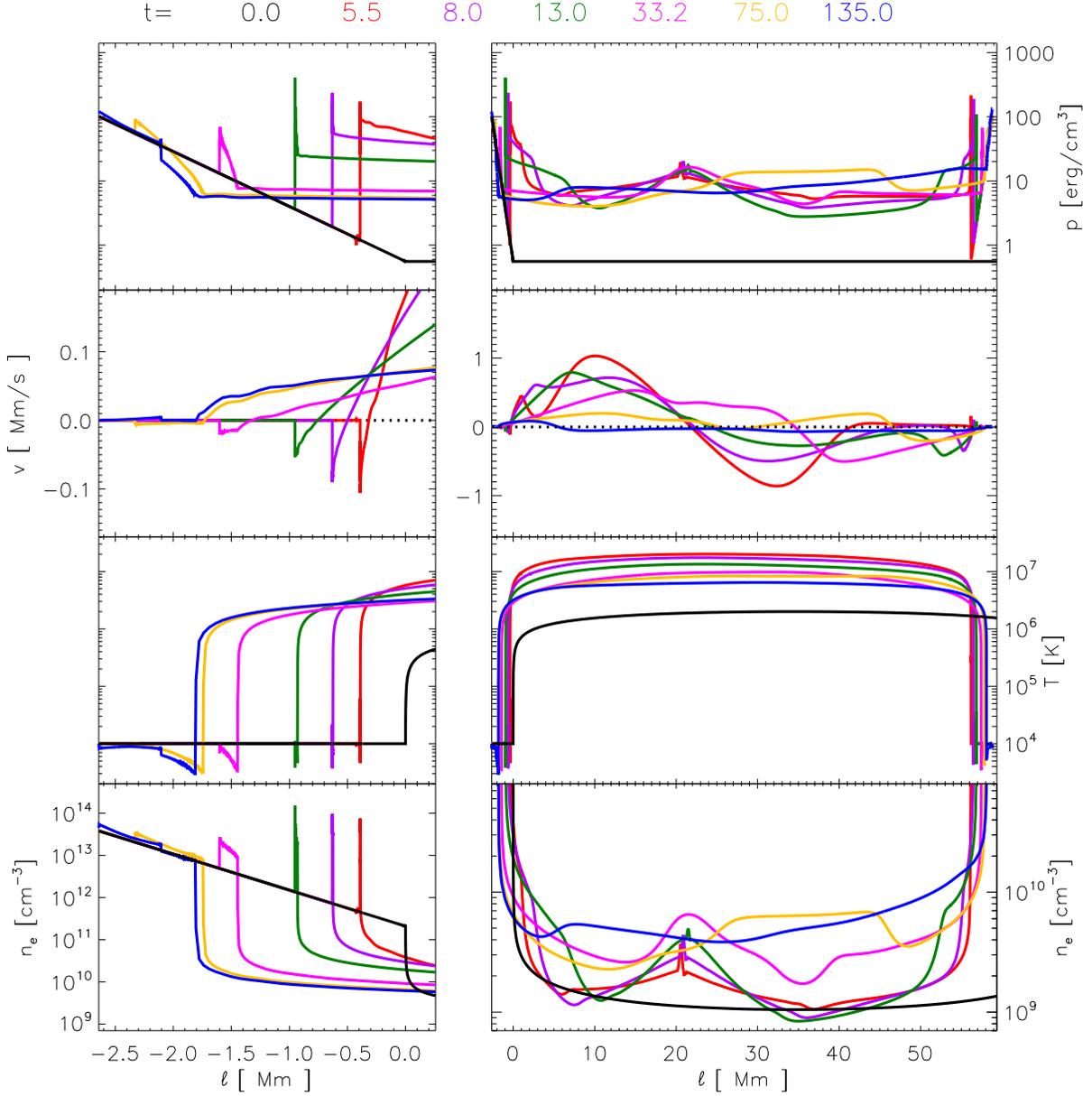}
\caption{Evolution of the TFT simulation after retraction has been halted.  Rows from top to bottom show pressure, velocity, temperature and electron number density at $t=0$ (black) and six later time in a color code given along the top.  All are plotted against length $\ell$, where $\ell=-2.6$ Mm is the bottom of the chromosphere so $\ell=0$ is the top in the initial tube (black).  The right column shows the entire straightened tube, out to $\ell=61.4$ Mm.  Since the tube was initially longer the right end, $\ell=74.5$, is not visible in the $t=0$ (black) plots.  The left column shows the evolution at the left footpoints.  The axes for velocity and density have been changed to better encompass the evaporative evolution.}
	\label{fig:bend45_evap}
\end{figure}

Figure \ref{fig:bend45_evap} shows the evolution of the straightened flux tube.  The evaporation, shown in the left column, resembles that in most other conductively driven models \citep[see][for a general discussion]{Longcope2014b}.  A condensation front\footnote{This is an historically rooted term which we feel compelled to use, in full knowledge that it has nothing to do with condensation is the conventional sense.} 
($v<0$) moves downward through the chromosphere.  It diminishes slightly as it goes, but by $t=135$ sec it has reflected from the fixed end of the tube.  To the right of the condensation is a rarefaction wave in which the velocity increases and density decreases rightward.  The rarefaction wave extends into the corona where it ends at an evaporation shock, or evaporation front, at which a density jump occurs.

The most novel aspects of this simulation are found in the coronal evolution, shown along the right column of 
\fig\ \ref{fig:bend45_evap}.  Instead of moving into ambient, stationary coronal material, the evaporation front encounters the persistent parallel flows created by the retraction.   At $t=5.5$ sec the left side side of the velocity profile shows a small bump from evaporation and a larger one from the RD.  By $t=8$ sec these have begun to merge and by $t=13$ sec there is a single velocity peak on the left.  The merging in the longer right leg is delayed by the later evaporation onset, but it is clearly complete by $t=33.3$ sec.  The central density pulse created by the 5 sec retraction remains visible until $t=13.0$ sec.   After this, the evaporation encounters it, enhancing the density further and moving the smoothed peak rightward.  This rightward motion is one clear result of the lack of perfect symmetry.  Symmetric evaporation fronts would have met in the center creating an even greater density enhancement which would not move.  Instead the density peak moves toward the farther footpoint, growing to $n_e\simeq1.5\times10^{10}\,{\rm cm}^{-3}$, some fifteen times above pre-flare level.

\begin{figure}[htp]
\plotone{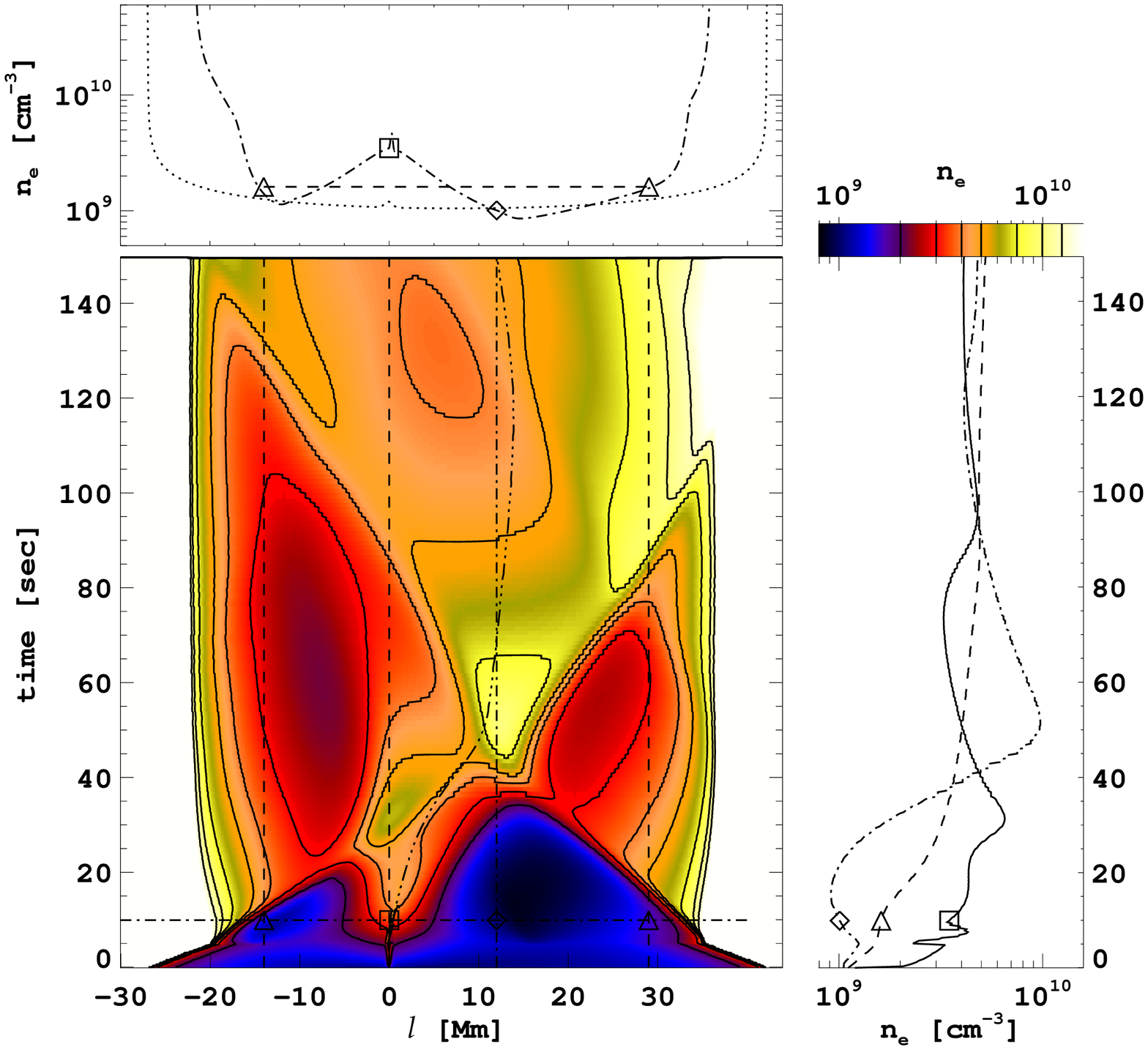}
\caption{The density evolution in the TFT simulation. The central panel shows density using a logarithmic color scale as a stack-plot, time {\em vs.} position.  Contours mark levels 
$n_e=[2,\,3,\,4,\,5,\,7.5,\,10,\,12.5]\times10^{9}\, {\rm cm^{-3}}$, which are also indicated on the color bar.
Snapshots are shown along a top panel, from both $t=0$ (dotted) and $t=10$ sec (dash-dot), the time indicated by a horizontal dash-dot line on the central panel.  Time-histories from select positions are plotted in the right panel.   Solid and dash-dot show positions $\ell=0$ (the point of reconnection) and $\ell=12$ Mm (the geometrical center).  The dashed curve is an average over $-14\,{\rm Mm}< \ell< 20\,{\rm Mm}$, shown by outer vertical dashed curves on the central panel.  Intersections with $t=10$ sec.\ are marked by triangles, plotted on all three panels.  The broken curve in the central panel shows the trajectory of the fluid element initially located at the reconnection site, $\ell=0$.}
	\label{fig:bend45_stack}
\end{figure}

A more complete picture of the density evolution is provided by a stack plot, \fig\ \ref{fig:bend45_stack}, showing density against both time (ordinate) and position (abscissa).  In order to combine the retraction phase ($t<5.0$ sec) with subsequent evolution we shift the origin, $\ell=0$, to the point where reconnection occurred.  The tube's left end is therefore initially at 
$\ell=-29.7$ Mm, and the right at $\ell=+44.8$ Mm.  As the tube retracts these points move inward, making the tube appear bounded by moving walls \citep{Longcope2010b,Longcope2011b,Brannon2014}.  This produces a wedge shape along the bottom of the stack plot.

The evaporation fronts appear in the central panel of \fig\ \ref{fig:bend45_stack} as wedges moving inward from each chromosphere.  The central density spike is a narrow tongue originating at $\ell=0$, and deflecting rightward at $t\simeq15$ sec, as the left evaporation front hits it.  The position of peak density is tracked through time (i.e.\ upward) by a broken curve.  The right evaporation front encounters the rightward-moving peak at $t\simeq40$ sec.  Thereafter it is possible to see a weaker transmitted version in the $n_e=5\times10^9\,{\rm cm}^{-3}$ contour.  More evident is the rightward continuation of the original peak, leaving density $n_e>7.5\times10^{9}\,{\rm cm}^{-3}$ on the right side at least until $t=150$ sec.

\begin{figure}[htp]
\epsscale{0.9}
\plotone{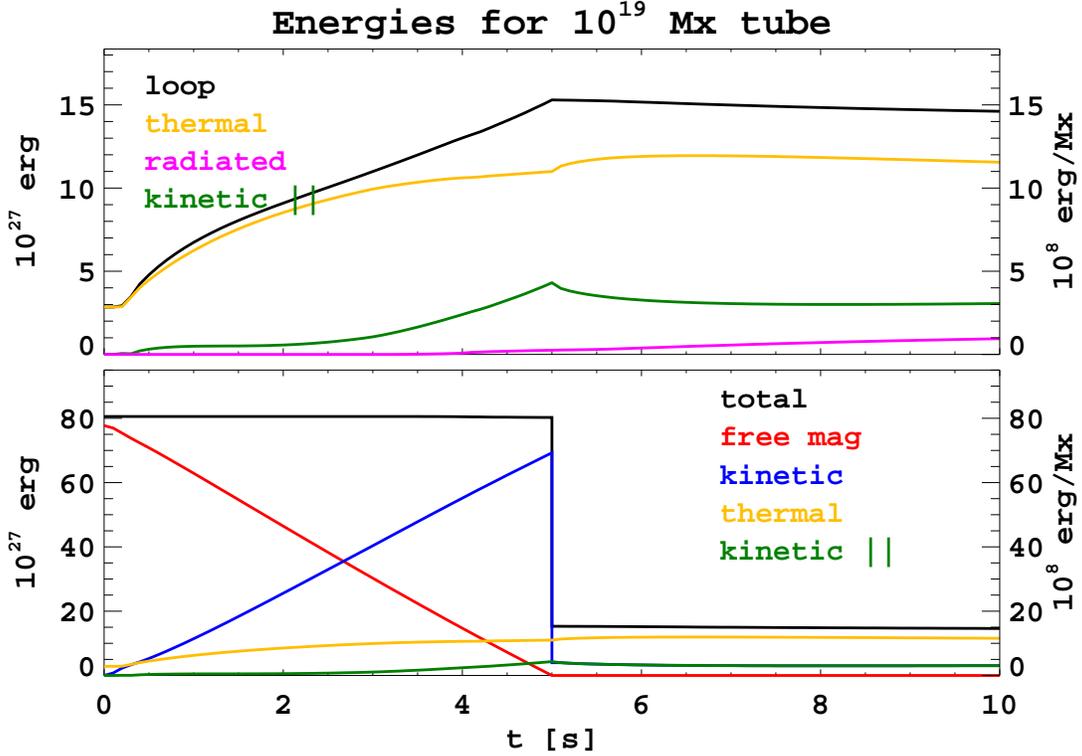}
\caption{Evolution of various energies during to first 10 sec.\ in the TFT model.  All energies are plotted in units of $10^8$ erg/Mx, which would be $10^{27}$ ergs for a $\Phi=10^{19}$ Mx flux tube.  The bottom panel shows the overall energetics, including free magnetic energy (red), kinetic energy (blue), thermal energy (orange) and their total (black).  The portion of kinetic energy in flow parallel to the tube is also plotted in green.  The upper panel shows thermal (orange) and parallel kinetic energy (green) on a zoomed scale.  The sum of these two, the loop energy, is plotted in black, and the energy lost to radiation is plotted in magenta.}
	\label{fig:erg45}
\end{figure}

The evolution of energies in this simulation are shown in \fig\ \ref{fig:erg45}.  Since we have fixed the magnetic field strength, the magnetic energy decreases only because the loop shortens by $\Delta L=13.1$ Mm, from $74.5$ to $61.4$ Mm.  This change in length produces a change in magnetic energy, per flux, of 
$\Delta W_m=B_0\Delta L/4\pi=7.8\times10^{9}\, {\rm erg/Mx}$.  The red curve in \fig\ \ref{fig:erg45} clearly falls from that level to zero over the 5.0 sec of tube retraction, thereby supplying the energy for the flare.  The magnetic energy is converted into $W_K=6.9\times10^{9}\, {\rm erg/Mx}$ of kinetic energy as well as adding 
$\Delta W_{th} = 8.1\times10^{8}\,{\rm erg/Mx}$ of thermal energy to the initial 
$W_{th,0}=2.8\times10^{8}\,{\rm erg/Mx}$.  An adiabatic 75:61 compression would have increased the thermal energy 
to $W_{th}=3.2\times10^{8}\,{\rm erg/Mx}$.  The thermal energy was increased by far more because the compression occurred at very high Mach number, basically in a shock.\footnote{In the TFT, as in conventional Petschek models, the ratio of thermal to kinetic energy is primarily a function of shear angle $\Delta\theta$ \citep{Longcope2009}.  In the limit of anti-parallel reconnection, $\Delta\theta=180^{\circ}$, the ratio achieves a well-known value of unity \citep{Priest2000}.}  During this entire time the gravitational energy has dropped by an insignificant $\Delta W_{g}=-0.93\times10^{8}\,{\rm erg/Mx}$, and a total of $0.25\times10^{8}\,{\rm erg/Mx}$ has been lost to radiation.

At $t=5.0$ sec the tube is straightened and all of the kinetic energy in flows perpendicular to the tube is discarded.  The kinetic energy then drops instantaneously to $W_{K,\parallel}= 4.3\times10^{8}\,{\rm erg/Mx}$.  The thermal energy is unchanged by the straightening, so the total energy after straightening is 
$W_{\rm loop}=15.3\times10^{8}\,{\rm erg/Mx}$, shown in black on the top panel of \fig\ \ref{fig:erg45}.  This represents about 20\% of the released magnetic energy, related to the fraction of kinetic energy downstream of the RDs directed parallel to the field, $v_x^2/v^2=\sin^2(\Delta\theta/4)=15\%$.  The remaining 80\%, converted into perpendicular motion, has been lost, presumably to fast magnetosonic waves, which are not trapped on the tube \citep{Longcope2012}.  Thereafter the kinetic energy is gradually thermalized, as a small portion of the sum is lost to radiation.

\section{The simulation with {\em ad hoc} heating}

How does the evolution described above compare with what would have happened if the same energy were introduced through the {\em ad hoc} heating term, $H(\ell,t)$, as in conventional models?  To answer this question we perform a simulation using the same code, but with a tube which is straight from the beginning.  To permit the clearest comparison we initialize this tube to have about the same length the TFT simulation had after straightening: $61.2$ Mm.  We designate 2.6 Mm at either end for the isothermal chromospheres, and thus initialize a tube with full coronal length $L_0=55.9$ Mm.  In order to give it roughly the same amount of material (i.e.\ electron column) as the initial TFT corona, we initialize this tube to have the same apex temperature, $T_{\rm max,0}=2.0\times10^6$.  As a result its pressure is 23\% higher than that of the 
initial TFT run.  The remaining properties are summarized in table \ref{tab:IC}.

Since heating in the TFT occurs within the corona, as it does in Petschek's model, we perform a simulation with 
{\em ad hoc} thermal heating, and refer to it as the {\em ad hoc} run.  A heating function, 
$H_{\rm fl}(\ell,t)$, is specified to deliver total energy equal to that ultimately delivered to the loop by retraction and subsequent straightening: $E_{\rm fl}=W_{\rm loop}-W_{\rm loop}(0)=12.4\times10^{8}\,{\rm erg/Mx}$.  It delivers that energy over the same $\tau=5.0$ sec interval, and to the same asymmetrically centered region.  To achieve this we define
\be
  H_{\rm fl}(\ell,t) ~=~ {4E_{\rm fl}B_0\over \tau\Delta\ell_{\rm h}}\,
  T(\,2t/\tau - 1\,)\, T\left( 2{\ell-\ell_{\rm h}\over\Delta\ell_{\rm h}} \right)
  	\label{eq:H}
\ee
where $T(x)$ is the normalized tent function
\be
  T(x) ~=~ \left\{ \begin{array}{lcl} 1-|x| &~~,~~& |x|\le1 \\ 0 &~~,~~&\hbox{otherwise} ~~. \end{array} \right.
\ee
The first tent function gives the heating a continuous time profile, like many previous simulations from the literature \citep{Cheng1983,Emslie1985,Bradshaw2003a}.   Heating in the TFT run, arising from viscous dissipation, is necessarily continuous although it probably rises more steeply than does the tent function.
The second tent function defines a heating region extending over the same tube portion covered by the RDs in the TFT run.  
This region extends  $\Delta\ell_{\rm h}=32.3$ Mm, and is centered asymmetrically
$\ell_{\rm h}=21.3$ Mm from the left chromosphere, which is $34.9$ Mm from the right.  The function $H_{\rm fl}(\ell,t)$ is added to the equilibrium, $H_{\rm eq}$, but since the former reaches $23.0\,{\rm erg\,cm^{-3}\,s^{-1}}$, 
it completely overwhelms the latter.

\begin{figure}[htp]
\epsscale{1.0}
\plotone{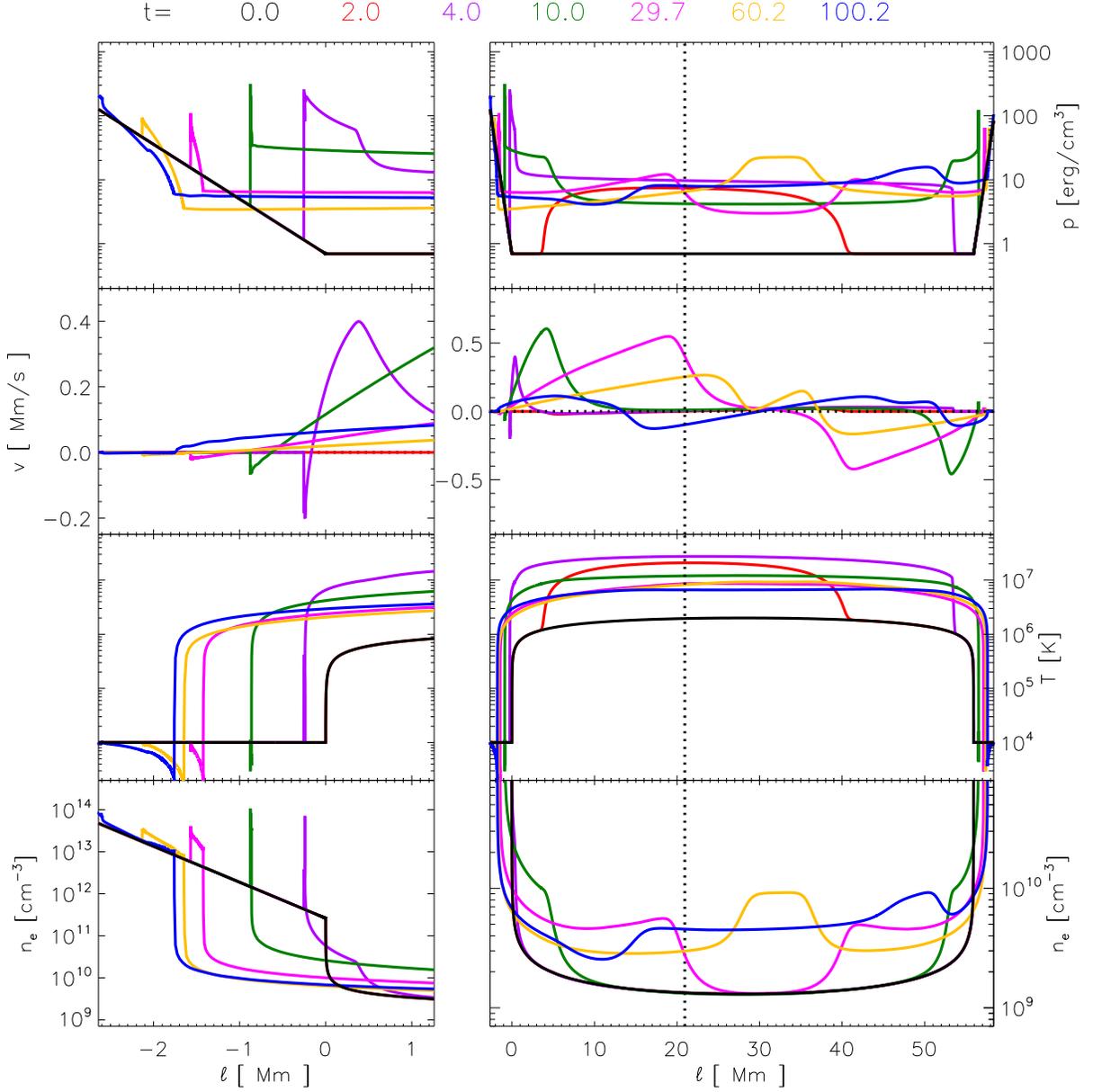}
\caption{The evolution of the {\em ad hoc} run plotted in the same format as \fig\ \ref{fig:bend45_evap}.  The dotted vertical line shows the center of the heating function, \eq\ (\ref{eq:H}).  Because there is no retraction phase, the times start earlier here, and the initial state ($t=0$, black) fits on the same axis as the others.}
	\label{fig:heat5_evap}
\end{figure}

The results of the {\em ad hoc} run, shown in \fig\ \ref{fig:heat5_evap}, resemble many from the previous literature.  The temperature is driven rapidly up, reaching $T=29$ MK by $t=3.1$ sec, with virtually no effect on the density.  Heat fronts move outward reaching each chromosphere in turn, and driving chromospheric evaporation.  The evaporation speed approaches the value, $v_e\simeq0.38(F/\rho_{co,0})^{1/3}=0.7$ Mm/s, predicted by \citet{Longcope2014b}, for the peak thermal flux, $F=E_{\rm fl}B_0/\tau=1.9\times10^{10}\,{\rm erg\,cm^{-2}\,s^{-1}}$.  Evaporation fronts are clearly seen propagating into an essentially undisturbed corona.  By $t=24.1$ they are poised to collide at a point slightly to the right of the geometric center, $\ell=28$ Mm.

\begin{figure}[htp]
\plotone{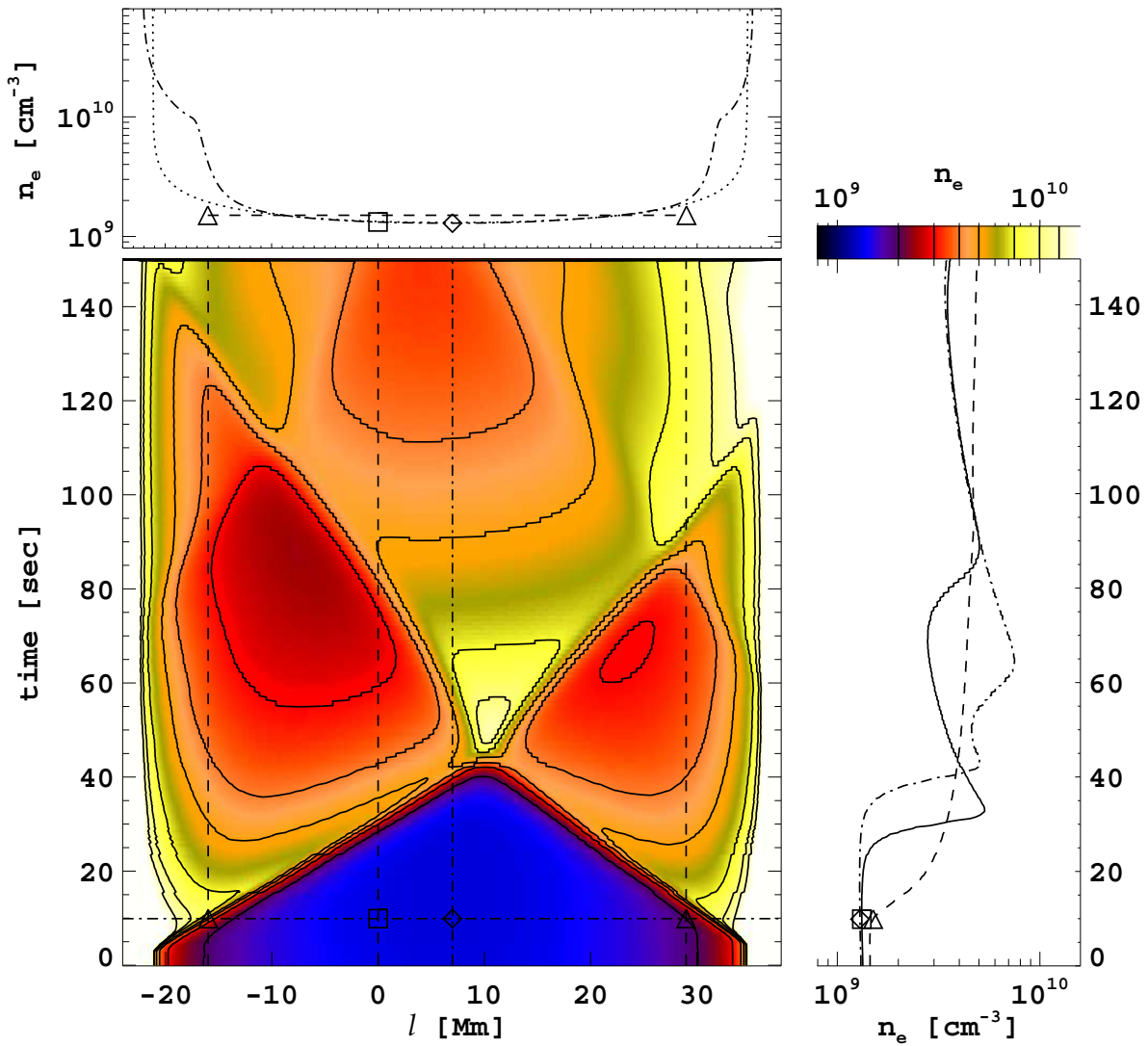}
\caption{Density evolution in the {\em ad hoc} run plotted in the same format as \fig\ \ref{fig:bend45_stack}.}
	\label{fig:heat5_stack}
\end{figure}

The true extent of the differences with the TFT run are clear in the density stack plot, \fig\ \ref{fig:heat5_stack}, where
we have again shifted the $\ell$ axis to put the center of the heating, $\ell_{\rm h}$, at the origin, and the geometric center at $\ell=6.8$ Mm.   The evaporation fronts form clear diagonal edges beneath which the corona has its pre-flare value, 
$n_e=1.3\times10^9\,{\rm cm^{-3}}$.  The fronts collide around $t=40$ sec just to the right of center, $\ell\simeq10$ Mm.  It seems that the heat front reaches the right side later, and is possibly slightly weaker, leading to a collision closer to that side.  In spite of this minor effect of asymmetry, the collision produces a virtually stationary enhancement, to
$n_e=1.2\times10^{10}\,{\rm cm^{-3}}$, which lasts only about 10 seconds.

\begin{figure}[htp]
\epsscale{0.9}
\plotone{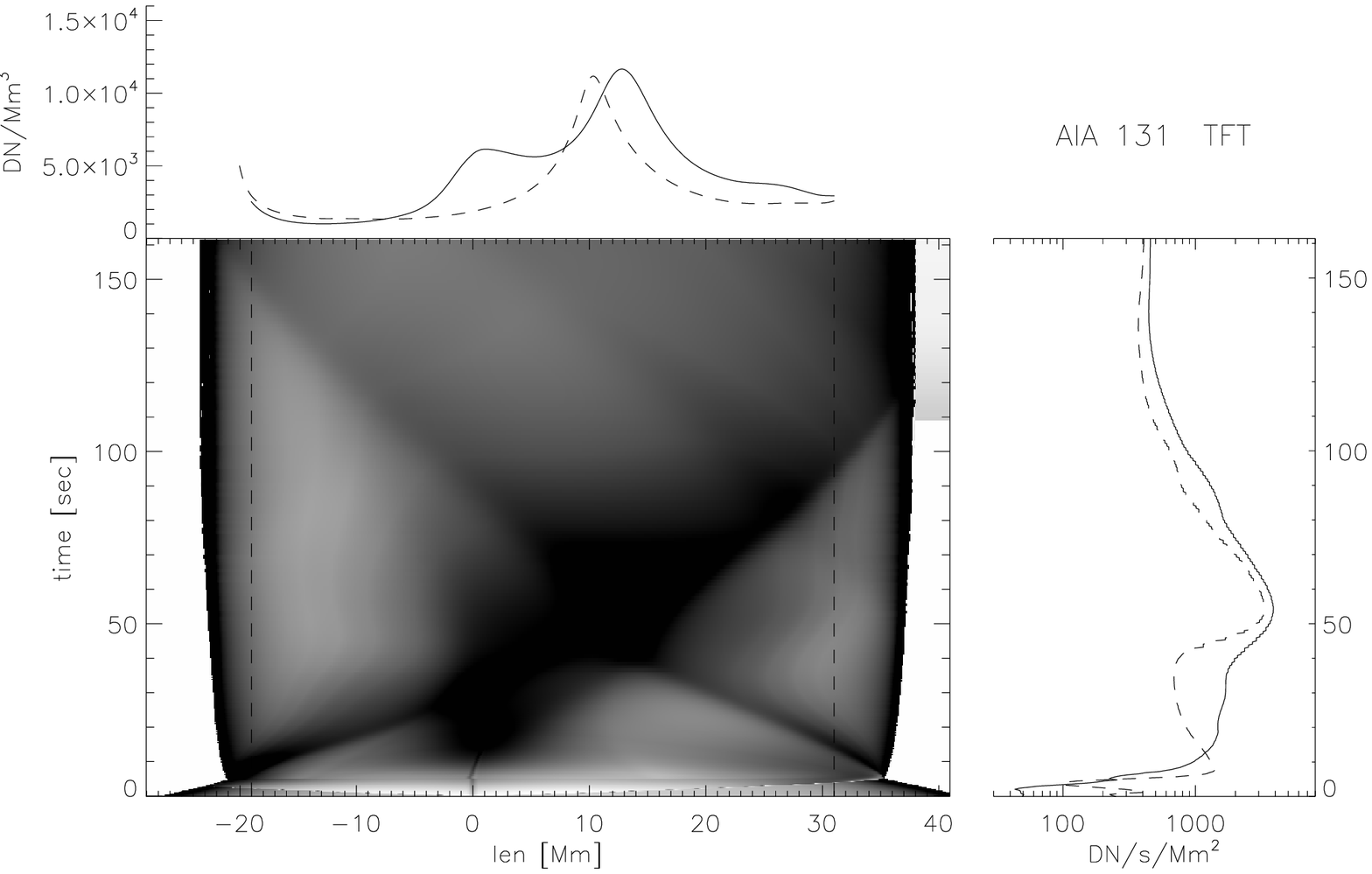}
\plotone{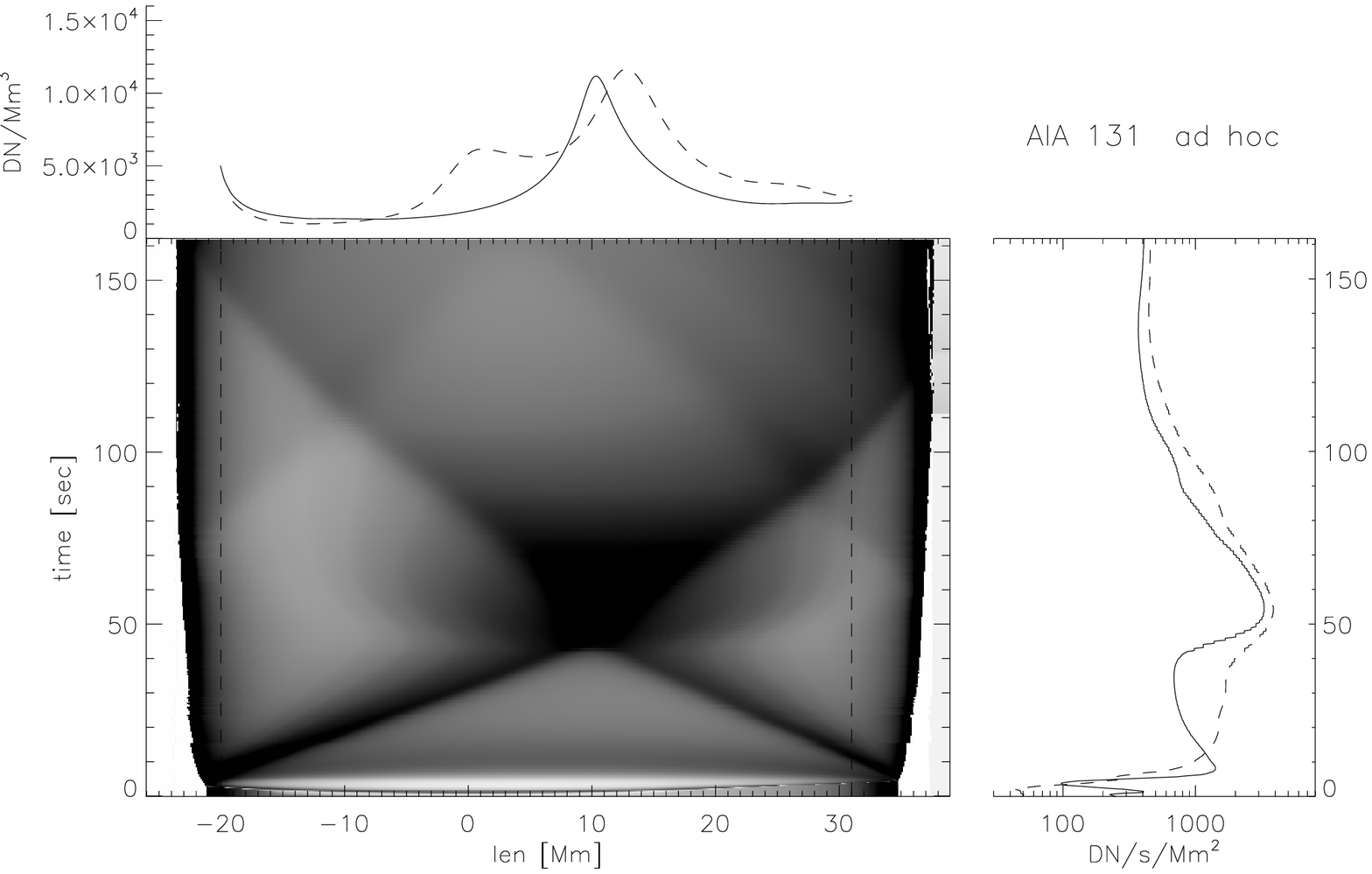}
\caption{Stack plots formed from synthetic AIA 131\AA\ emission for the TFT run (top set) and {\em ad hoc} run (bottom set).  The logarithmic, reverse grey-scale (black is high) in each central panel shows emission, and the dashed vertical lines define the coronal portion.   The solid curves along the right show its integral over the corona; the dashed curve shows the other version for comparison.  The curve across the top shows the total time integral for a given position.}
	\label{fig:AIA131}
\end{figure}

It seems that the two models of energy supply lead to different density evolution.  These differences are illustrated well using synthesized EUV emission.  Figure \ref{fig:AIA131} shows AIA 131\AA\ emission synthesized from the temperature response returned by the \verb!aia_get_reponse! function in \verb!ssw! \citep{Freeland1998}.   The bandpass is sensitive to coronal plasma primarily in Fe {\sc xxi} emission which peaks at $T=10$ MK \citep{Lemen2012} under the assumption of ionization equilibrium made deriving the response function.  
The results, plotted as stack plots similar to \figs\ \ref{fig:bend45_stack} and \ref{fig:heat5_stack}, reflect the density structures in those figures, but only in the hottest portions.  The TFT has a more pronounced asymmetry, with the integrated emission (top curves) coming predominantly from the right half.  The {\em ad hoc} run shows evaporation fronts far more clearly, since there is no outflow-generated density structure to impede them.  These exhibit a slight asymmetry, but the dominant feature is the first collision very near the center.  The TFT run also produces more hot coronal emission during the initial 50 seconds.  This is the result of simultaneous heating and compression by the slow shocks.  The {\em ad hoc} run lacks that compression.

In spite of their differing histories of coronal densities, the two simulations converge to very similar longer-term evolution.  After $t\simeq60$ sec the integrated 131 \AA\ light curves run in close parallel, with the TFT model being slightly brighter.  At this point in each model, the pressure has become roughly uniform and flows have mostly subsided.  The loops are therefore in approximate mechanical equilibrium as they continue to cool.  Synthetic emission from cooler lines (not shown) show little of the early, dynamic phase, and thus do not differ markedly between the runs.  It seems that since the same energy was added to each loop, they reach similar equilibria, and follow similar cooling profiles.

\begin{figure}[htp]
\epsscale{1.0}
\plotone{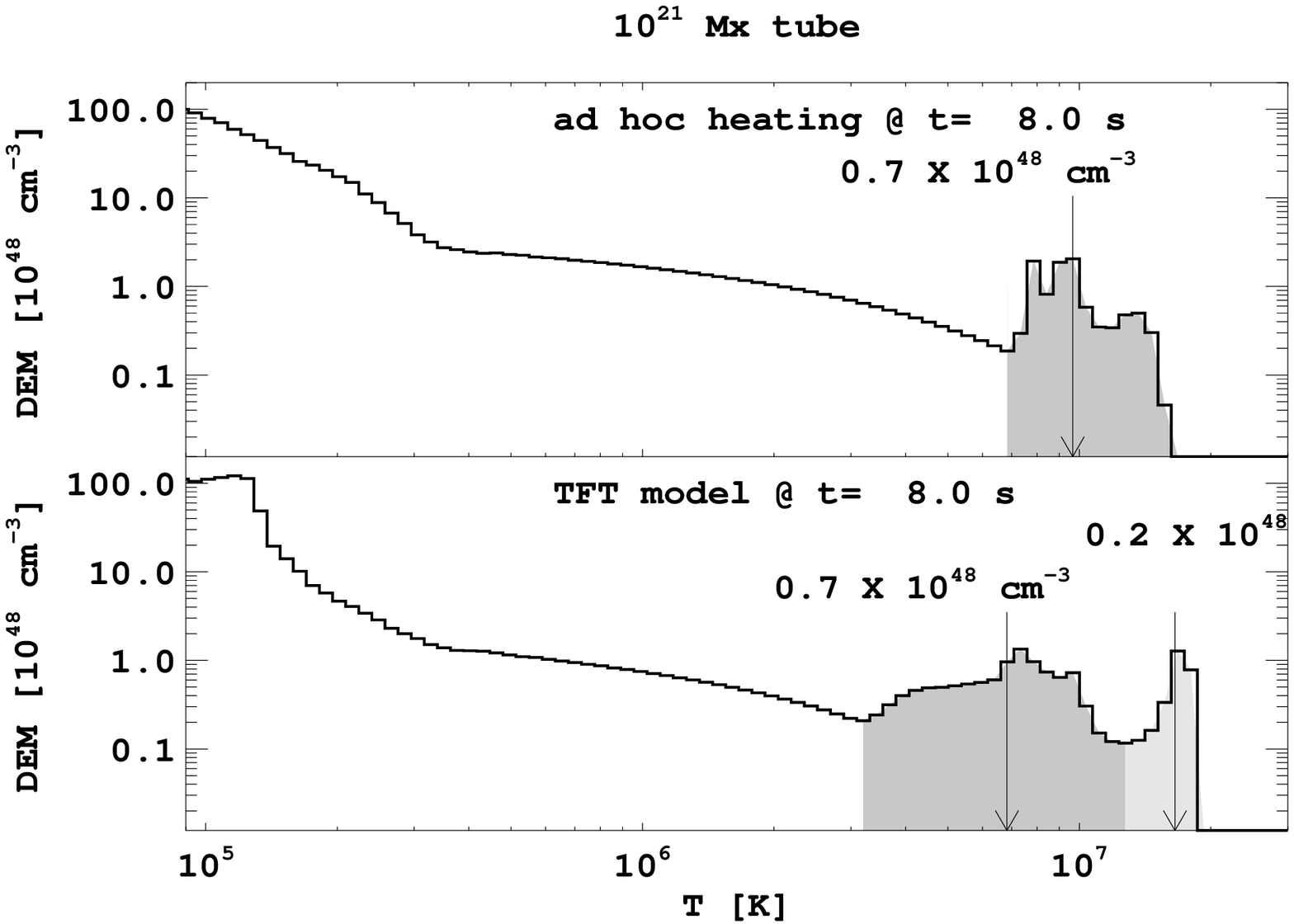}
\caption{Differential emission measures (DEMs) formed by averaging each run over the 2.0 sec interval centered at 8.0 sec.  In order to make results comparable to full-flare values, the DEM is multiplied by a flux of $10^{21}$ Mx, comparable to an entire active region.  Portions of each DEM are identified with evaporation and heating, and colored with dark and light grey shading respectively.  The total emission from each component is labelled, and its centroid indicated with an arrow.}
	\label{fig:DEM}
\end{figure}

The heating and compression of the TFT model produces hot loop-top concentration more effectively than the {\em ad hoc} model.  This is evident by comparing differential emission measures (DEMs) at the same time in each simulation.  The DEMs shown in \fig\ \ref{fig:DEM} result from 2 sec integration of the runs at a time after heating has ended, and evaporation has been active.  Both cases have evaporated plasma, shaded in darker grey, 
of roughly the same emission measure, $0.7\times10^{48}\,{\rm cm}^{-3}$, 
but that from the TFT run has slightly lower mean temperature, $\bar{T}=7$ MK {\em vs.} $\bar{T}=10$ MK for the 
{\em ad hoc} run.  This plasma started very cool, and now has temperatures distributed from about $0.5$ MK to over 10 MK.   The TFT run also has a distinct {\em hot} component, $EM\simeq0.2\times10^{48}\,{\rm cm}^{-3}$, at 
$\bar{T}=17$ MK, resulting from the retraction-generated shocks.  This has persisted even after the retraction has ceased.

\begin{figure}[htp]
\epsscale{0.85}
\plotone{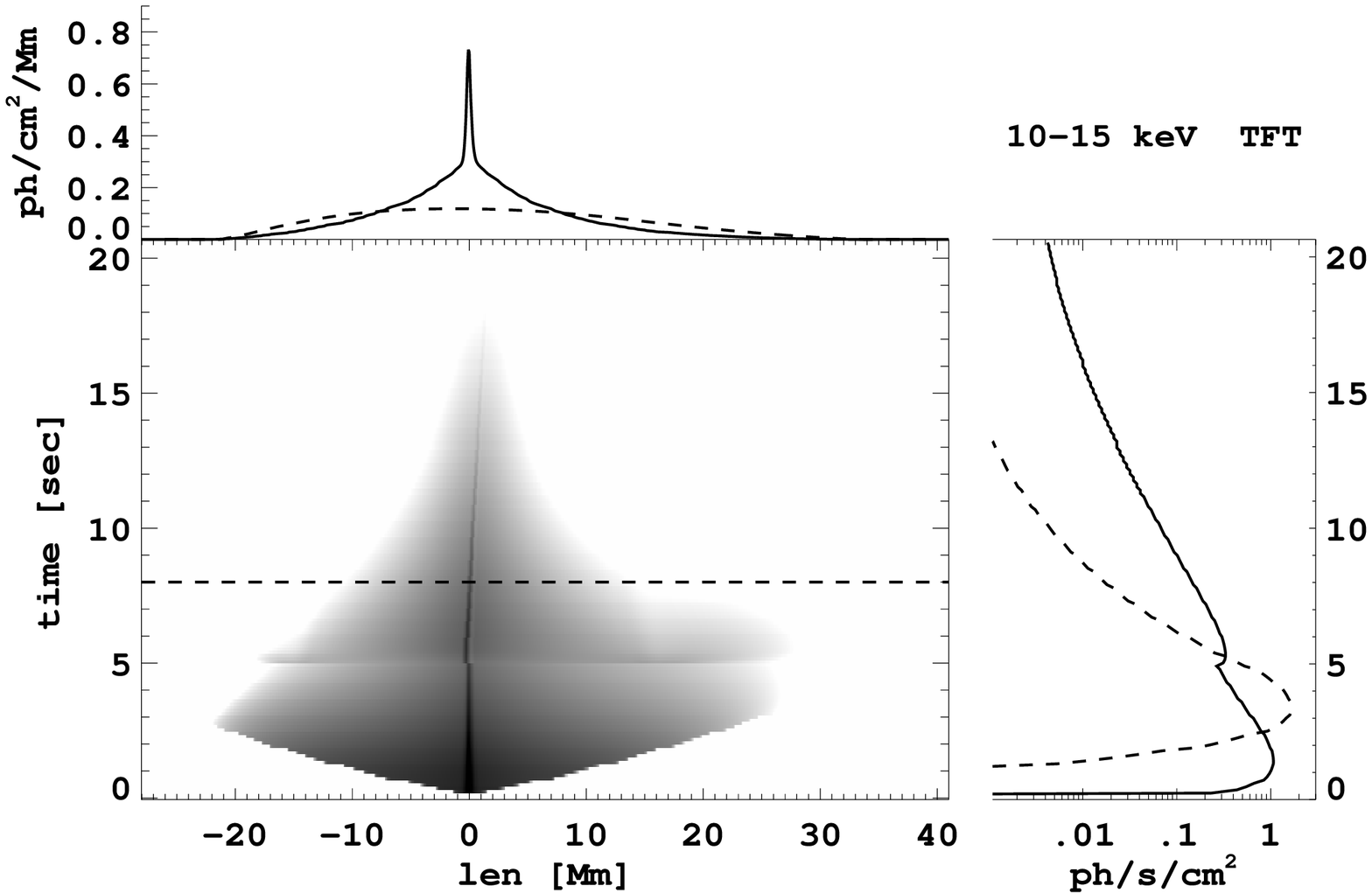}
\plotone{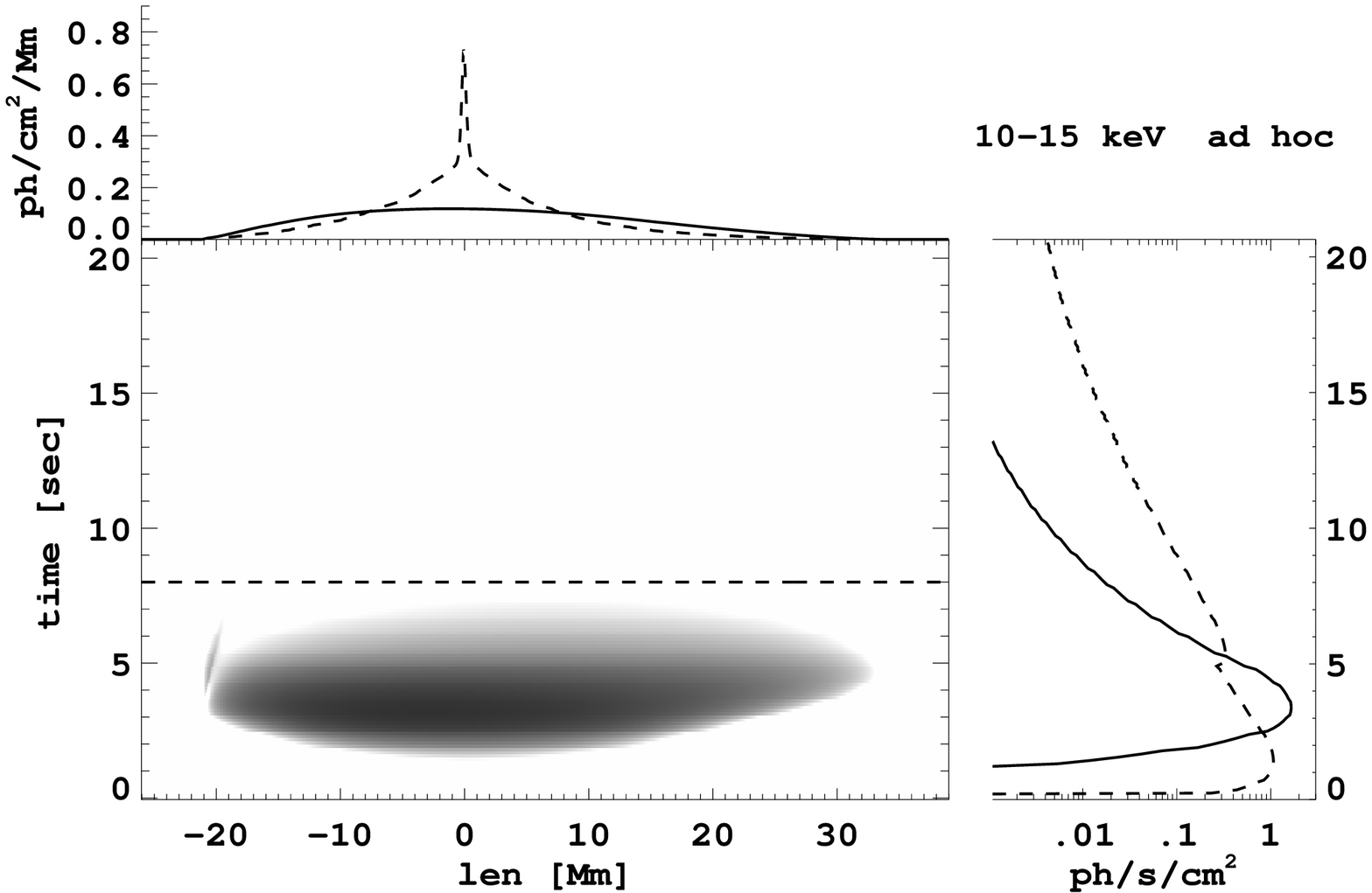}
\caption{Stack plots formed from synthetic HXR images, in a format similar to \fig\ \ref{fig:AIA131}, but for times running only to $t=20$ sec.  The center panels show, in logarithmic, inverse grey-scale, emission in the 10--15 keV range.  Horizontal dashed lines show the time, $t=8.0$ sec, at which the DEM from \fig\ \ref{fig:DEM} were made.  The right panels show integrated photon flux from a single tube of $\Phi=10^{19}$ Mx.  The dashed curve corresponds to the other model, to facilitate comparison.  The top panels show the time-integrated emission from each point along the loop.}
	\label{fig:HXR}
\end{figure}

The hot density concentration of the TFT run gives rise to observable loop-top\footnote{We use this term interchangeably with the observationally distinct term {\em above-the-loop-top} source.  Our single-loop simulation cannot predict its location relative to cooler loops heated earlier.  It therefore cannot distinguish between loop-top and above-the-loop-top sources.} hard X-ray (HXR) emission, shown in \fig\ \ref{fig:HXR}.  Photon flux at 1 AU in the 10--15 keV range, due to thermal bremsstrahlung \citep[\eq\ {[}5.14b{]} of][]{Rybicki1979}, is computed for each segment of the loop.  These are multiplied by the cross sectional area of an elemental tube, $\Phi=10^{19}$ Mx, of which many dozen will be emitting at any time in a typical flare.  The TFT plot (top) shows a bright (dark in the figure's inverse color table) streak of HXR emission concentrated at the reconnection site, $\ell=0$, persisting to $t\simeq15$ sec, long after retraction ends ($t=5$ sec).  Fainter emission expands outward with the flux-limited conduction fronts, only to become squelched when they encounter the chromosphere, $t\simeq3$ and 4 seconds on the left and right respectively.  These two components are evident in the time-integrated curves along the top.  The stack plot  from the {\em ad hoc} run (bottom) shows only the broad, diffuse emission reflecting the time history of the heating function far more clearly than it shows the conduction fronts.  There is no loop-top source in the HXR in this model.   Both loops have cooled so much by the time the evaporations fronts collide ($t\simeq40$ sec) that no HXR emission is visible.

\begin{figure}[htp]
\epsscale{1.0}
\plotone{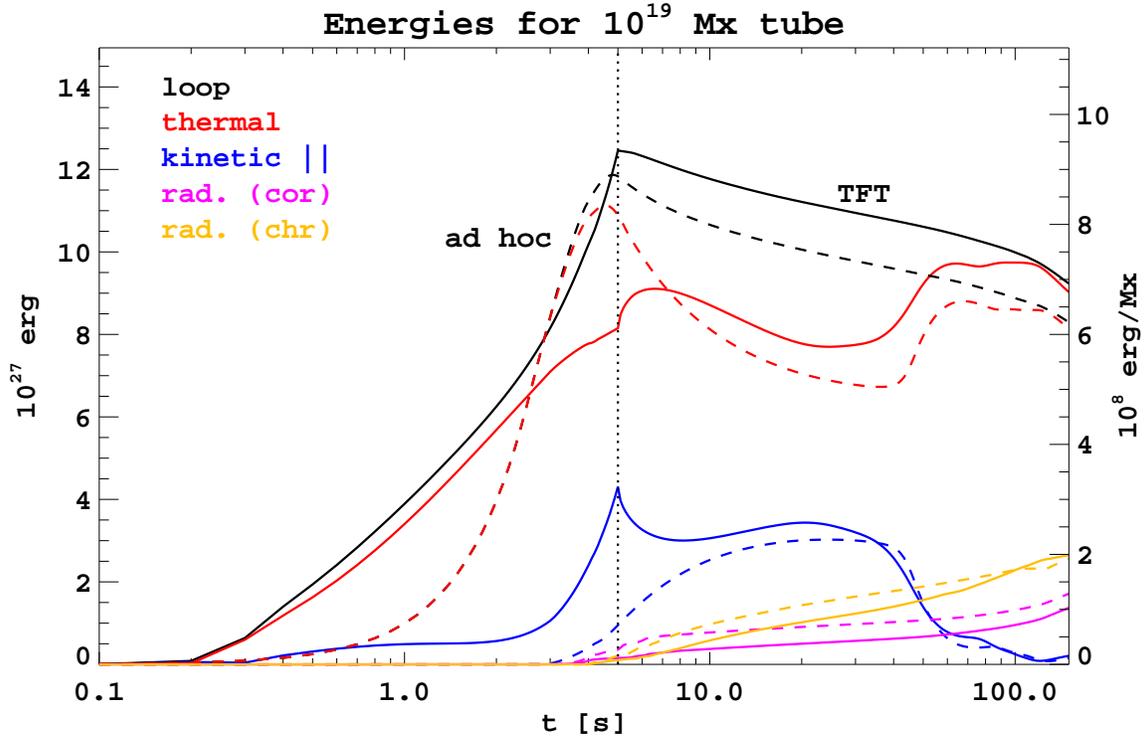}
\caption{The evolution of energies for the two runs, plotted against a logarithmic time axis.  The dotted vertical line at $t=5.0$ sec shows the end of the energy release phase in each run.  Solid and dashed curves show energies for the TFT and {\em ad hoc} runs respectively, according to a color table on the upper left.  The magenta lines show the cumulative loss from radiation by plasma above $10^5$, referred to here as coronal radiation.  Orange curves shows the cumulative losses from cooler plasma, called chromospheric losses.}
	\label{fig:erg_cmp}
\end{figure}

The final area for comparison is energetic evolution.  Figure \ref{fig:erg_cmp} shows extended time series of the energies of each run.  A logarithmic time axis is used to best show all the phases of evolution.  The total energy (``loop energy'' which is thermal plus parallel kinetic) in each case rises rapidly during the 5.0 sec energy release.  The continuous, tent-shaped time profile of $H_{\rm fl}(\ell,t)$, in \eq\ (\ref{eq:H}), leads to a more parabolic rise in the {\em ad hoc} run.   A run with flat-topped profile (not shown) delivering the same total energy as the tent, over the same time interval, produces a more linear rise.  Since it has a lower peak heating rate, $H_{\rm fl}$, it reaches a peak temperature ($26$ MK) lower than the {\em ad hoc} simulation shown here.  By the end of the heating phase the flat-topped run has achieved virtually the same point as the {\em ad hoc} run and evolves virtually identically thereafter.

The most significant difference between the TFT and {\em ad hoc} runs is evident in the interplay between thermal and kinetic energy, shown by red and blue curves respectively.  All flow in the {\em ad hoc} run comes from thermally-driven evaporation.  The kinetic energy can only increase at the expense of thermal energy, and the red and blue dashed curves clearly move out of phase.  In the TFT model, by contrast, the flux tube retraction generates significant kinetic energy (blue solid curve) from the beginning even as the thermal energy rises.  After the retraction has ceased the kinetic energy is thermalized causing the thermal energy to continue rising.  It seems that by $t=10$ sec, flows are being sustained by evaporation in both runs, and their kinetic energies evolve together.  By $t=70$ sec both runs have entered a cooling phase where kinetic energy has almost vanished and thermal energy decreases only through radiation.  The {\em ad hoc} run has slightly less energy at late times, primarily due to the loss of $\simeq0.5\times10^8\,{\rm erg/Mx}$ by coronal
radiation during the heating phase.  Its coronal losses continue to slightly exceed those of the TFT run, causing it to fall beneath the TFT run by about $0.7\times10^8\,{\rm erg/Mx}$ as of $t=150$ sec.

\section{Discussion}

The foregoing comparisons have revealed several differences between flare models whose energization emulates Petschek reconnection model 
and those in which energy is delivered {\em ad hoc}.  In our TFT model, as in Petschek's, magnetic energy is converted first to bulk kinetic energy by the Lorentz force, and thereafter to heat by shocks.  The {\em ad hoc} model adds energy directly as heat.  As a consequence, the TFT case retains more energy in flows than does the {\em ad hoc} simulation.  All flows in the latter are driven by evaporation, meaning that kinetic energy arises from thermal energy rather than {\em vice versa}.  In an extension of this logic, once the flows have been dissipated, both models follow nearly identical evolution.


The most evident difference between the models was observed in their early loop-top density.  Each model showed density enhancement from the collision of evaporation flows, beginning about 40 sec.\ after energy release began -- the time required for $\sim700$ km/sec flows to traverse the 30 Mm half-length.  The post-reconnection retraction of the TFT run produced its own loop-top density concentration, and did so almost immediately.  This concentration persisted well after the retraction, until the evaporation flow could reach it.  The result was a far more pronounced and long-lasting loop-top source, and one that could be observed in high-temperature EUV and HXR emission, due to its appearance during the flare's hottest phase.  In the {\em ad hoc} run, energy was added to a static loop, so this early, hot density concentration was absent.

The loop-top concentration of the TFT model offers a sufficient barrier to the evaporation flows to produce EUV loops with a distinctly asymmetric appearance.  This asymmetry is most obvious in the time-integrated 131\AA\ curves, plotted along the tops of the stack plots of \fig\ \ref{fig:AIA131}.  The TFT model has emission primarily from the right ($\ell>0$) half of the loop. In a genuine EUV image, this asymmetry would give the impression of a post-flare {\em half-loop}, of which there are reports in the literature \citep{Tsuneta1996,Forbes1996,Guidoni2015}.  \citet{Forbes1996} proposed that the half-loop appearance was an artifact of the viewing angle.  More recently, \citet{Guidoni2015} used stereoscopic observations to rule this out, and proposed instead that a shock at the loop top was impeding evaporation from filling half the loop.  The TFT simulation done here supports that interpretation, although only at the hotter lines such as 131\AA.  Further investigation will be needed to understand how these features could persist into the cooler lines, as they are observed to do \citep{Guidoni2015}.

The TFT is used here to mimic the mechanism by which Petshcek's reconnection model energizes plasma through reconnection.  Under this interpretation, the TFT's loop-top concentration represents a manifestation of Petschek's slow magnetosonic shocks.  If a continuous flux transfer were synthesized from individual loop retractions, the result would have the appearance of a steady Petschek outflow jet \citep{Longcope2010b}.  The concentration in the TFT persists well past the end of the retraction phase, placing it at the termination of the outflow jet.  
This is also where fast magnetosonic terminations shocks (FMTSs) are generally sought \citep{Forbes1986c,Magara1996,Chen1999,Sui2003}, but the artificial straightening used to halt the retraction was designed to omit compression which would be associated with fast magntosonic waves.  Thus the stationary density concentration we observe cannot be a FMTS.  Instead it is a remnant of the slow mode shock, confined by the momentum of the parallel flows left over from the RDs.  \citet{Longcope2010} predicted this feature and estimated its persistence time using analytical arguments; the current model corroborates that rather crude estimate.  To the extent that the TFT model mimics other aspects of reconnection in multi-dimensional models, we propose that previous MHD simulations be revisited with the aim of determining if loop-top density concentrations exhibited in them are actually FMTSs, or persistent slow mode residuals, or a combination of the two.



Our TFT model is inspired by the simple, yet widely-studied, reconnection model of Petschek.  This origin led directly to several of its less realistic features including Alfv\'enic outflows and a lack of non-thermal electrons.  When reconnection outflows are observed, they rarely have speeds more than a fraction of the Alfv\'en speed \citep{Savage2011}.  This  fact is at odds with most numerical reconnection simulations which do have outflows speeds comparable to the local Alfv\'en speed \citep{Magara1996,Yokoyama1997,Chen1999}.  This discrepancy must be explored from a more general perspective before we can hope to incorporate reconnection into more realistic flare model.  

Along a similar vein, global reconnection models, such as Petschek's, are based on fluid equations and lack non-thermal particles.   Numerous investigations have explored the possibility that particular features of this model, such as magnetic islands, turbulence, or stochastic waves, might give rise to a population of non-thermal particles \citep{Miller1997,Petrosian2004,Drake2006}.  These investigations are promising, but have not yet progressed to the point that they form a complete self-consistent model of flare energy release through reconnection.  Once such a model is formulated in the traditional multi-dimensional context, it will be a relatively straightforward task to incorporate it into a TFT model and produce flares loops.


Lacking a non-thermal electron population both our models relied on thermal conduction to transport energy along the field lines.  We modified the thermal conductivity to keep the heat flux below the free streaming limit (see \eq\ [\ref{eq:kappa}]).  We chose a limiting factor, $\xi=1$, larger than that used in many previous investigations, $\xi\simeq1/6$ \citep{Emslie1985}.  In a brief survey we observed that smaller values led to higher flare temperatures, smaller loop-top density peaks, and larger pressure jumps at the conduction front.  The latter pressure jump became so large for values of $\xi$ below $1/2$, that the Petschek reconnection appeared to turn ``inside out'': the thermal fronts began to resemble slow mode shocks, but located 
{\em outside} the RDs.  This had other unusual consequences which warrant future investigation.  We are not aware of any reconnection simulations using flux-limited, field-aligned thermal conductivity, where similar structures could have been previously observed.  For present purposes we therefore opted to use the value, $\xi=1$, in order to keep heat flux below the free-streaming limit, while at same time producing reconnection with conventional structure.



Our simulations made other approximations of a less fundamental nature.  In particular, we adopted the model of \citet{Longcope2009} and \citet{Guidoni2010} in which the retracting flux tube is confined within a planar current sheet, which is equivalent to taking the reconnection opening angle $\Delta\phi\to0$ in \fig\ \ref{fig:geom}.  Since the dynamics of the retracting flux is dominated by the magnetic tension of the initially bent flux tube, the most important factor is the bend angle $\Delta\theta$ of the initial condition.  Having taken the opening angle $\Delta\phi\to0$ we equated this initial angle with the shear angle of the equilibrium current sheet.  Had we chosen, instead, to consider a broader outflow wedge, the initial bend angle would be slightly larger than the equilibrium shear angle.  Since the subsequent dynamics depend only on this initial angle our TFT run simulation would be largely unaffected by the adoption of a finite opening angle except that our run, with $\Delta\theta=90^{\circ}$, would actually correspond to reconnection at a shear angle slightly smaller than $90^{\circ}$.  This adjustment is clearly small enough to have been undetected in previous comparisons between the TFT model and steady, two-dimensional reconnection models with finite $\Delta\phi$ \citep{Longcope2010,Longcope2011b}.

The TFT model is, however, simply a proxy for reconnection, and it is useful to the extent that it includes elements in common with fast magnetic reconnection.  Like a reconnection model, magnetic energy is released in the TFT through retraction of field lines.  In both models this leads to a high-speed jet, and shocks where plasma is both compressed and heated.  By including these elements in a flare loop model we have shed some light into how reconnection might interact with other elements of a flare, such as chromospheric evaporation.  While only a proxy, the TFT offers more realism than the majority of flare loop models, which introduce heat through {\em ad hoc} source terms, and thus lack those reconnection elements we have investigated.  Since the TFT is not a full model of reconnection, a definitive test of our findings must await a multi-dimensional MHD simulation.

The foregoing work has made a careful comparison between a single simulation of each type, using a single set of flare parameters.  We chose parameters characteristic of a modest flare, $B_0=75$, $\Delta\theta=90^{\circ}$, $L_{\rm fin}=62$ Mm, and $T_{\rm max,0}=2\times10^6$ K (table \ref{tab:IC} gives the complete list).  Reconnecting a total flux $\Delta\Phi=10^{21}$ Mx composed of such tubes would release $8\times10^{30}$ ergs, of which $1.5\times10^{30}$ ergs would ultimately produce flare-related radiation.  Other cases with different parameters, not reported here, show qualitatively similar behavior, with predicable quantitative differences.  Longer loops evolve more slowly, shorter loops more quickly.  Higher field strength, $B_0$, or shear angle, $\Delta\theta$, produce more intense retraction shocks with higher temperature \citep{Longcope2009}.  This may be the route to producing loop-top HXR sources more substantial than that seen in our modest flare.  Observations of super-hot sources, $T\sim40$ MK, are believed to arise from reconnection at field strengths often $200$ G or more 
\citep{Longcope2010,Caspi2010,Caspi2014}.

\acknowledgements

This work was supported by a grant from NASA's Heliophysics Supporting Research program.


\end{document}